\documentclass{jfm}
\usepackage{graphicx}
\usepackage{epstopdf, epsfig}
\usepackage{color}
\usepackage{lineno,hyperref}
\usepackage{amsmath}

\definecolor{pinegreen}{rgb}{0.0, 0.47, 0.44}
\definecolor{green(html/cssgreen)}{rgb}{0.0, 0.5, 0.0}

\newcommand{\eref}[1]{(\ref{#1})}
\newcommand{\pref}[1]{(\ref{#1})}
\newcommand{\fref}[1]{Figure \ref{#1}}
\newcommand{\sref}[1]{\S \ref{#1}}
\newcommand{\aref}[1]{Appendix \ref{#1}}

\shorttitle{Wall-to-wall optimal transport}
\shortauthor{A. N. Souza, I. Tobasco and C. R. Doering}

\title{Wall-to-wall optimal transport in two dimensions}

\author{Andre N. Souza\aff{1}
  \corresp{\email{andrenogueirasouza@gmail.com}},
  Ian Tobasco\aff{2}
 \and Charles R. Doering\aff{2,3,4}}

\affiliation{
\aff{1}Department of Earth, Atmospheric, and Planetary Sciences, Massachusetts Institute of Technology, Cambridge, MA 02139
\aff{2}Department of Mathematics, University of Michigan, Ann Arbor, MI 48109-1043
\aff{3}Center for the Study of Complex Systems, University of Michigan, Ann Arbor, MI 48109-1042
\aff{4}Department of Physics, University of Michigan, Ann Arbor, MI 48109-1140}

\begin{document}

\maketitle

\begin{abstract}
Gradient ascent methods are developed to compute incompressible flows that maximize heat transport between two  isothermal no-slip parallel walls.
Parameterizing the magnitude of velocity fields by a P\'eclet number $\text{Pe}$ proportional to their root-mean-square rate-of-strain, the schemes are applied to compute two-dimensional flows optimizing convective enhancement of diffusive heat transfer, i.e., the Nusselt number $\text{Nu}$ up to $\text{Pe} \approx 10^5$.
The resulting transport exhibits a change of scaling from $\text{Nu}-1 \sim \text{Pe}^{2}$ for $\text{Pe} < 10$ in the linear regime to $\text{Nu} \sim \text{Pe}^{0.54}$ for $\text{Pe} > 10^3$.
Optimal fields are observed to be approximately separable, i.e., products of functions of the wall-parallel and wall-normal coordinates.
Analysis employing a separable ansatz yields a conditional upper bound $\lesssim \text{Pe}^{6/11} = \text{Pe}^{0.\overline{54}}$ as $\text{Pe} \rightarrow \infty$ similar to the computationally achieved scaling.
Implications for heat transfer in buoyancy-driven Rayleigh-B\'enard convection are discussed.

\end{abstract}

\begin{keywords}
variational methds, maximal transport, Rayleigh-B\'enard Convection
\end{keywords}

\section{Introduction}
%\iffalse  \fi 

Heat transfer via fluid advection is a critical component of atmospheric, oceanographic, geophysical, and astrophysical dynamics, as well as being the basis of cooling systems in engineering applications.
Numerous studies on how to design systems that achieve enhanced heat transfer by either manipulation of domain geometry or through the discovery of suitable flow structures have recently appeared in the literature \cite{srikanth2017, Silas_improved_cooling, fmarcotte2018, osakaheat2,osakaheat}.
A particularly fruitful approach to discovering flow structures was first introduced by \cite{HassanzadehChiniDoering2014} where it was formulated via an optimal control approach.

The original motivation for \cite{HassanzadehChiniDoering2014} was to develop a new tool for obtaining upper bounds on thermal transport by buoyancy-driven flows, e.g., for Rayleigh-B\'enard convection.
The analysis and derivation of upper bounds on transport properties plays a prominent role in expanding our knowledge of fundamental fluid dynamics and serves a complimentary role to other methods of inquiry, i.e., direct numerical simulations of the underlying equations of motion, invoking closure models to determine statistical properties, or postulating phenomenological models of turbulent transport.
The first proof of upper bounds on heat transfer by Rayleigh-B\'enard convection was achieved in \cite{Howard63}.
The complementary ``background method" was subsequently introduced in \cite{DC1996}.
Both approaches leverage certain bulk integral constraints derived from the equations of motion and yield bounds which apply to a strictly larger class of flows. It remains unknown whether the resulting bounds are realizable by buoyancy driven flows.

Unlike those previous approaches, the wall-to-wall transport problem introduced in \cite{HassanzadehChiniDoering2014} fully enforces the advection-diffusion equation for the temperature field pointwise in space and time.
Admissible incompressible advecting flow fields do not (necessarily) satisfy an equation of motion, but are instead subject to a bulk integral intensity constraint, i.e., fixed finite magnitude of energy or enstrophy, and suitable boundary conditions.
This allows consideration of the following question: Amongst all possible incompressible flow fields subject to a fixed intensity constraint and relevant boundary conditions, which ones maximize thermal transport?

As usual, we model heat transport with the advection-diffusion equation
\begin{align}
\partial_t T + \nabla \cdot \left( \mathbf{u} T - \kappa \nabla T \right) &= 0
\end{align}
where the coefficient $\kappa$ is the thermal diffusivity.
The two-dimensional spatial domain is $\Omega = [0,L_x) \times [0,L_z]$ and the temperature field $T$ is periodic in the horizontal $x$ direction, ``hot" on the bottom boundary where $T(z=0) = T_0$ and ``cool" on the top boundary where $T(z=L_z) = T_1$ with $T_0 > T_1$.
The advecting flow field $\mathbf{u} = u_1 \hat{x}+ u_3 \hat{y}$ is divergence-free ($\nabla \cdot \mathbf{u} = 0$) with no penetration through the boundaries, i.e., $u_3(z=0)=0=u_3(z=L_z)$.
In this paper the velocity is restricted to satisfy no-slip boundary conditions on the top and bottom boundaries, $u_1(z=0)=0=u_1(z=L_z)$.
Both components are periodic in the horizontal $x$ direction. Other boundary conditions can be handled using similar methods.
Initial ($t=0$) data for the temperature field is provided to formally pose the evolution problem for $t>0$. 
Using units $L_z$ and $L_z^2 / \kappa$ for space and time and changing the temperature $T \rightarrow  \frac{T- T_0}{T_1 - T_0}$ transforms the system to 
\begin{align}
\label{advection_diffusion}
\partial_{t} T + \nabla \cdot \left( \mathbf{u} T -  \nabla T \right) = 0,
\end{align}
with $x \in [0, \Gamma )$ where $\Gamma = L_x / L_z$, $z \in [0,1]$, and with $T(z=0) = 1$ and $T(z=1) = 0$. We consider \eref{advection_diffusion} henceforth.

Given a flow field $\mathbf{u}$ the non-dimensional measure of thermal transport is the space- and long time-average of the convective heat flux in the vertical direction, i.e., the Nusselt number
\begin{align}
\text{Nu} \{ \mathbf{u} \} &= \langle u_3 T - \partial_z T  \rangle 
\end{align}
where $\langle \cdot \rangle$ denotes the space-time average
\begin{align}
\langle f \rangle &\equiv \limsup_{\tau \rightarrow \infty} \frac{1}{\tau \Gamma }  \int_0^\tau \int_0^1 \int_0^{\Gamma}   f(x,z,t) dx dz dt .
\end{align}
The boundary conditions imply unit mean conductive heat flux, i.e.,  $\langle -\partial_z T \rangle = 1$.
 
In this work, the goal is to determine the largest possible value of $\text{Nu}$ as a function of $\mathbf{u}$, which we parameterize by a P\'eclet number defined as the root-mean-square vorticity (the square root of the mean enstrophy density) $\text{Pe} \equiv \langle  | \nabla \times \mathbf{u} |^2 \rangle^{1/2}$.
% \begin{align}
%\text{Pe}^2 &=  \langle  | \nabla \times \mathbf{u} |^2 \rangle.
%\end{align}
Incompressibility and the boundary conditions imply that $\text{Pe}$ is simply related to the norm of $\nabla \mathbf{u}$ and the mean-square rate of strain:
\begin{align}
\text{Pe}^2 &=  \langle  | \nabla \times \mathbf{u} |^2 \rangle =  \langle  | \nabla \mathbf{u} |^2 \rangle = 2 \langle  | (\nabla \mathbf{u})_{\text{sym}} |^2 \rangle  \equiv | \nabla u_1 |^2 + | \nabla u_2 |^2.
\end{align}
The optimal wall-to-wall transport problem is then to maximize $\text{Nu}$ as a function of $\text{Pe}$.
Explicitly, we take on the task to
\begin{align}
\label{primitive}
&\text{Maximise} \text{ }  \langle u_3 T \rangle \text{ } \text{subject to } \\
\nonumber
&\partial_t T+ \mathbf{u} \cdot \nabla T = \Delta T, \\
& \nabla \cdot \mathbf{u}=0, \ \langle | \nabla \mathbf{u} |^2 \rangle = \text{Pe}^2, \\
&\text{and the boundary conditions}.
\end{align}

Note in the wall-to-wall problem the velocity field $\mathbf{u}$ is not required to satisfy conservation of momentum. Nevertheless, in \cite{HassanzadehChiniDoering2014} its solution was shown to inform bounds on buoyancy-driven driven transport; indeed, the original motivation for introducing the wall-to-wall problem was to find new upper bounds on the Nusselt number in Rayleigh-B\'enard convection modeled by the Boussinesq equations 
\begin{align}
\label{navier_stokes}
&\partial_t \mathbf{u} + \mathbf{u} \cdot \nabla \mathbf{u} + \nabla p = \text{Pr} \Delta \mathbf{u} + \text{Pr} \, \text{Ra} \,T   \, \hat{z}, \\
\label{incompressible}
&\nabla \cdot \mathbf{u} = 0, \text{ and}\\
&\partial_t T +  \mathbf{u} \cdot \nabla T = \Delta T.
\end{align}
Here, the Prandtl number $Pr$ and the Rayleigh number $\text{Ra}$ are given by
\begin{align}
{Pr} = \nu / \kappa \text{ \ and \ } \text{Ra} = \frac{ \alpha g (T_1 - T_0) (L_z)^3 }{\kappa \nu }
\end{align}
where $\nu$ is the kinematic viscosity, $g$ is the acceleration due to gravity, and $\alpha$ is the thermal expansion coefficient.
%The non-dimensionalization used to obtain this form of the Boussinesq equations is the same as \eref{nondim} and the boundary conditions on $T$ and $\mathbf{u}$ are the same as for the optimal wall-to-wall problem above.
The challenge for Rayleigh-B\'enard is to derive upper bounds on the convective heat transport of the form $\text{Nu} - 1 \leq f( \text{Pr}, \text{Ra} )$ that hold for all solutions of the Boussinesq system.
The connection between the P\'eclet number $\text{Pe}$ and the Rayleigh number $\text{Ra}$ is obtained by dotting $\mathbf{u}$ into \pref{navier_stokes} and averaging over space and time (utilising integration by parts with the given boundary conditions and \pref{incompressible}) to obtain the identity $\text{Pe}^2 = \text{Ra} \left(\text{Nu}-1 \right)$.
As a result, any upper bound for the wall-to-wall problem implies an upper bound for convective heat transport among solutions of the Boussinesq equations.  
For example, an upper bound of the form $\text{Nu} -1 \leq  c \text{Pe}^{\beta}$ implies an upper bound of the form $\text{Nu} -1 \leq c^{\frac{2}{2-\beta}}\text{Ra}^{\frac{\beta}{2-\beta}}$. 

The optimal wall-to-wall transport problem is evidently non-convex --- there may be many local maxima and global extrema --- and only by evaluating the global maximum are we assured of an upper bound for Rayleigh-B\'enard.
Nevertheless, to make progress, we seek local maxima numerically in this paper and discuss the resulting flows. These flows are of interest in their own right as mechanisms to significantly enhance heat transport, e.g., as targets for control.  %The best class of solutions one can solve for numerically are local maxima.
%The resulting flows are of interest in their own right as they can achieve significantly enhanced heat transport over pure conduction. %Hence our interest in this study is to examine flow structures that enhance heat transport.
%\cblue{such as structures of engineering interest for the design of heat transfer enhancing flows \citep{osakaheat2,osakaheat,fmarcotte2018}?}.

%\cblue{The related 3D computations in \cite{osakaheat2} exemplifies the many local maxima that can exist in higher dimensions. In addition to every 2D solution they have found flow structures that exhibit nearly optimal heat transport, $\text{Nu} \sim \text{Pe}^{2/3}$, that bifurcate from the 2D solutions at $\text{Pe} \approx 79.2$. The optimal 3D solutions are markedly different from the 2D computations and seem to point out fundamental differences between 3D and 2D flow fields. In 2D it is still an open question if one can find flows that exhibit a similar optimal scaling; however Tobasco et al. \cite{tobasco2017, tobasco2018} have rigorously shown that there exist flow fields with a nearly optimal scaling,  ${\cal O} \{\frac{Pe^{2 / 3} }{\log ( \text{Pe} )^{4/3}} \}$ as $\text{Pe} \rightarrow \infty$.  }

The rest of this paper is organized as follows.
We first introduce Lagrange multipliers to implement the constraints of the wall-to-wall optimal transport problem and examine the structure of the resulting functional in \sref{f_a_c}.
From insights gained on manipulations of the functional, we develop time-stepping algorithms in \sref{grad_flow} to solve the Euler-Lagrange equations.
Solutions to the saddle point conditions and resulting transport scalings are presented for time-independent two-dimensional flow fields with no-slip boundaries in \sref{results}.
Upon investigation of the numerical solutions we see that the fields are to a very high degree separable, i.e., the computed stream functions satisfy $\psi(x,z) \approx \phi(x)\Psi(z)$ and similarly for the other fields.
This numerical observation motivates an analytic examination of upper bounds on the wall-to-wall problem with an additional separable ansatz --- which is apparently \textit{almost} satisfied by solutions of the wall-to-wall Euler-Lagrange equations --- in \sref{tensor_product_upper_bounds} leading to conditional upper bounds on heat transport of the form $\text{Nu} \lesssim  \text{Pe}^{6/11}$ or, in terms of Rayleigh-B\'enard, $\text{Nu} \lesssim \text{Ra}^{3/8}$.

 Along the way, we discuss the relationship of the wall-to-wall optimal transport problem to the background method of \cite{DC1996} in \sref{background_method}, and to the Howard-Busse-Malkus problem \cite{malkus_convection,Howard63,Busse69} in \sref{hbm_prob}.
The perspective developed in those sections inspires the design of a time-stepping algorithm for computing optimal flows, similar to that in \cite{Baole2015} for computing optimal background fields.
Our methods of temporal and spatial discretization are described, respectively, in \sref{time_evo} and \aref{disc_details}.
In particular we find in \sref{ascent_methods} that evolving the equations 
 \begin{align}
\partial_\tau \xi &= \Delta \xi - \mathbf{u} \cdot \nabla \eta  + u_3 \\
0 &= -\Delta \eta - \mathbf{u} \cdot \nabla \xi  \\
\partial_\tau \mathbf{u} &= \mu \Delta \mathbf{u}  - \xi \nabla \eta + \xi \hat{z} + \frac{1}{2} \nabla p  \\
0 &= \nabla \cdot \mathbf{u} 
 \end{align}
forward in pseudo-time  $\tau$ (for a fixed constant $\mu$) subject to homogeneous boundary conditions for $\eta$ and $\xi$ and no-slip boundary conditions for $\mathbf{u}$ yields local maxima of the wall-to-wall problem.
One of the many benefits of the algorithms described here is that optimal flow fields may be computed for other geometries, e.g., cylinders, given suitable Poisson and Stokes' equations solvers.

 \section{Theory}
\label{f_a_c}
In this section we utilise Lagrange mutlipliers to rewrite the wall-to-wall optimisation problem as one of finding saddle points of a certain unconstrained functional $\mathcal{F}$.
We then describe various manipulations that can be performed on $\mathcal{F}$ involving its maximization or minimization or both in \sref{eta_variation}-\sref{w2w_chara}. This leads us to a direct comparison between the background method and the wall-to-wall problem in \sref{background_method}. In particular we conjecture that there exists a duality gap between the two problems in \sref{poly_examp} and provide a simple polynomial example to illustrate why one should expect the problems to be distinct. Additionally we note the connection to the Howard-Busse-Malkus problem in \sref{hbm_prob}. All of these considerations aid us in the development of numerical algorithms for producing candidate optimizers in \sref{grad_flow}, and in the proof of our conditional upper bounds in \sref{tensor_product_upper_bounds}.
%We also identify here of these results we will obtain the connection between the wall-to-wall problem and the background method.

We begin by introducing a new temperature variable $\theta = T - (1-z)$ and rewrite the advection diffusion equation as
\begin{align}
\partial_t \theta + \mathbf{u} \cdot \nabla \theta = \Delta \theta + u_3.
\end{align}
Next we introduce Lagrange multipliers $\mu$ (a positive real number), $p(x,z,t)$ and $\varphi(x,z,t)$ and the unconstrained functional
\begin{align}
\label{fund}
\mathcal{F}&= \left\langle u_3 \theta + \varphi\left(-\partial_t \theta - \mathbf{u}\cdot \nabla \theta + \Delta \theta +u_3 \right) + \mu\left(\text{Pe}^2 -  | \nabla \mathbf{u} |^2\right) + \nabla p \cdot \textbf{u} \right\rangle,
\end{align}
the saddle points of which are candidates for the maximization problem \eref{primitive}. The variables $\varphi$ and $p$ come equipped with their natural boundary conditions. Namely, we impose periodic and homogeneous boundary conditions in the $x$ and $z$ directions respectively for $\varphi$, and the usual implicit boundary conditions for $p$.

Given initial data one could search for time-dependent optimal flow fields of the wall-to-wall problem, but we restrict ourselves to time-independent flow fields for a number of reasons.
First, steady fields are far easier to compute numerically and time dependence greatly expands the scope and range of our current endeavor.
Second, in the context of simpler models such as the Lorenz equations (with a ``heat transport'' functional analogous to the Nusselt number), time-dependence was found to never increase transport \cite{SouzaDoeringLorenz, Souza201526}.
Third, preliminary attempts at computing time-dependent flow fields for the wall-to-wall problem have yielded \textit{essentially} time-independent results, suggesting that time-dependence may not play a role in significantly enhancing heat transport. More precisely, taking the initial condition of the temperature field to be the conductive state $1-z$ we found that the result of the time-dependent optimization was to move the conductive state into a (locally) optimal steady state, and to hold it there.
Finally, for time-independent flows we are guaranteed that maximisers exist (and that the functional $\mathcal{F}$ is differentiable) while as of now there is no such assurance for time-dependent flows. 

With these considerations in mind, from this point on we focus on the time-independent functional
\begin{align}
\label{ffund}
\mathcal{F} &= \left\langle u_3 \theta + \varphi\left( \mathbf{u}\cdot \nabla \theta + \Delta \theta +u_3 \right) + \mu\left(\text{Pe}^2 -  | \nabla \mathbf{u} |^2\right) + \nabla p \cdot \textbf{u} \right\rangle
 \end{align}
where now the brackets $\langle \cdot \rangle$ are understood to give the spatial average only.
We seek the saddle points of \eref{ffund} and for this it will be useful to consider alternative coordinate systems or ``constraint manifolds'' that pass through these.
Many of the manipulations introduced below extend naturally to time-dependent and/or stress-free flow fields.

The first manipulation we perform is to change variables by $\theta = \xi + \eta$ and $\varphi = \xi - \eta$, following the ``symmetrization method" described in \cite{tobasco2017} and \cite{tobasco2018}.
 Various integrations by parts yield the functional
\begin{align}
\label{cfund}
\mathcal{S}&= \left\langle |\nabla \eta |^2 + 2\xi \mathbf{u}\cdot\left(  \hat{z} - \nabla \eta  \right)- |\nabla \xi |^2 + \mu\left(\text{Pe}^2 - | \nabla \mathbf{u}|^2 \right) + \nabla p \cdot \textbf{u} \right\rangle .
\end{align}
The advantage of the new $(\xi, \eta)$ coordinates lies in exposing the underlying geometry of the wall-to-wall problem. Indeed, the functional $\mathcal{S}$ is convex with respect to $\eta$ and concave with respect to $\mathbf{u}$ or $\xi$ when the other is held fixed, whereas the functional $\mathcal{F}$ fails to have such properties. It is a bit like choosing to study the polynomial expression $x^2 - y^2$ instead of $(x+y)(x-y)$.

As a warm up to the manipulations that follow, let us consider this simple example a bit more and remark on how one might search for the saddle points of $s(x,y) = x^2 - y^2$.
Gradient ascent/descent procedures are problematic on their own, but can be successfully combined with constraints picking out certain curves.
For example, the curve $y=0$ passes through the saddle point $(0,0)$ and the resulting function $s(x,0) = x^2$ can be minimized by gradient descent.
Thinking procedurally, this particular constraint curve is found by taking the derivative of $s$ with respect to $y$ and setting the result equal to zero, i.e., $\frac{\partial s}{\partial y} = 2y = 0$. 

Returning to the functional $\mathcal{S}$, we proceed in \sref{eta_variation} and \sref{xi_u_variation} to derive various constraint manifolds that pass through its saddle points. We do so by setting the variations of $\mathcal{S}$ with respect to $\eta$, $\xi$, or $\mathbf{u}$ equal to zero and relating these to optimisations of $\mathcal{S}$. Then, in \sref{w2w_chara} and \sref{background_method} we show how the wall-to-wall optimal transport problem and the background method arise from two such optimisation procedures, thereby producing insights into the relationship between the two.

\subsection{Variations with respect to $\eta$}
\label{eta_variation}
We start by taking the variation of $\mathcal{S}$ with respect to $\eta$ and setting it equal to zero. This results in the Euler-Lagrange equations
\begin{align}
\label{opt_eta}
\Delta \eta &=  \mathbf{u} \cdot \nabla \xi
\end{align}
for $\eta$. Substituting this back into $\mathcal{S}$ results in the constrained functional
\begin{align}
\label{inf_eta_func}
 \mathcal{S}_\eta &\equiv \left\langle 2 u_3 \xi -|\nabla \Delta^{-1} \left(\mathbf{u} \cdot \nabla \xi  \right) |^2 - |\nabla \xi |^2 + \mu\left(\text{Pe}^2 - | \nabla \mathbf{u} |^2 \right) + \nabla p \cdot \textbf{u} \right\rangle.
\end{align}
Stated differently, constraining the variable $\eta$ to satisfy \eref{opt_eta} preserves the saddles of \eref{cfund} and yields 
\begin{align}
\mathcal{S}_\eta &= \min_\eta \mathcal{S}.
\end{align}
%\cblue{This may seem rather strange given that in the end we want to maximize heat transport, but recall that we are interested computing saddle points of $\mathcal{F}$ and $\mathcal{S}$ rather than the original wall-to-wall problem. (remove??)}

%A related functional is utilized in the context of effective diffusivity in convection enhanced diffusion \cite{doi:10.1137/S0036139992236785}. 

\subsection{Variations with respect to $\xi$ and $\mathbf{u}$}
\label{xi_u_variation}
Next, we take the variation of $\mathcal{S}$ with respect to the variable $\xi$ and set it equal to zero. This yields
\begin{align}
\Delta \xi &=  \mathbf{u} \cdot \nabla \eta - u_3
\end{align}
and after substituting it back into $\mathcal{S}$ we find 
\begin{align}
\label{sup_xi_func}
\mathcal{S}_\xi &=  \left\langle |\nabla \eta |^2 + |\nabla \Delta^{-1} \left( \mathbf{u}\cdot\left(   \hat{z}-\nabla \eta  \right)\right) |^2 + \mu\left(\text{Pe}^2 - | \nabla \mathbf{u} |^2\right) + \nabla p \cdot \textbf{u} \right\rangle .
\end{align}
As before, the same functional is obtained by maximizing $\mathcal{S}$ in $\xi$, i.e., 
\begin{align*}
\mathcal{S}_\xi &= \max_{\xi} \mathcal{S}.
\end{align*}

Finally, we take variations over all incompressible flow fields $\mathbf{u}$. This produces the constrained functional
\begin{align}
\mathcal{S}_{\mathbf{u}} &= \left\langle |\nabla \eta |^2  - |\nabla \xi |^2 + \mu \text{Pe}^2 + \frac{1}{\mu } | \nabla S^{-1} \left( \xi  \hat{z} - \xi \nabla \eta  \right)|^2  \right\rangle
\end{align}
where $S^{-1}\left( \xi  \hat{z} - \xi \nabla \eta  \right)$ denotes the unique flow field $\mathbf{u}$ solving
\begin{align}
\mu \Delta \mathbf{u} &= \xi \nabla \eta - \xi \hat{z} - \frac{1}{2}\nabla p , \\
\nabla \cdot \mathbf{u} &= 0.
\end{align}
Note that 
\begin{align*}
\mathcal{S}_{\mathbf{u}} &= \max_{\mathbf{u}}\mathcal{S}.
\end{align*}

These observations serve as a starting point in \sref{tensor_product_upper_bounds} for establishing upper bounds on transport and motivate our choice of numerical methods in \sref{grad_flow}. But first, let us see how the wall-to-wall problem comes out of these manipulations.

\subsection{Finding the wall-to-wall problem}
\label{w2w_chara}

Consider the structure of the $S_\eta = \min_\eta \mathcal{S}$ and $S_\xi = \max_{\xi} \mathcal{S}$ functionals for a fixed incompressible velocity field $\mathbf{u}$ with enstrophy $\langle | \nabla \mathbf{u} |^2 \rangle = \text{Pe}^2$.
The functional $S_\eta$ is concave in $\xi$; likewise $S_\xi$ is convex in $\eta$.
Thus finding the maximum of the former with respect to $\xi$, or the minimum of the latter with respect to $\eta$, is equivalent to enforcing their Euler-Lagrange equations. 

The Euler-Lagrange equation for $S_\eta$ in $\xi$ is 
\begin{align}
\label{xi_fo}
\Delta \xi &=  \mathbf{u} \cdot \nabla \Delta^{-1}\left( \mathbf{u} \cdot \nabla \xi \right) - u_3.
\end{align}
Similarly for $S_\xi$ in $\eta$ we have
\begin{align}
\label{eta_fo}
\Delta \eta &=  \mathbf{u} \cdot \nabla \Delta^{-1} \left( \mathbf{u} \cdot \nabla \eta - u_3\right) .
\end{align}
These can be written more concisely as the system
\begin{align}
\Delta \eta &=  \mathbf{u} \cdot \nabla \xi \\
\Delta \xi &=  \mathbf{u} \cdot \nabla \eta - u_3,
\end{align}
gotten by setting
\begin{align}
\eta &= \Delta^{-1} \left( \mathbf{u} \cdot \nabla \xi \right) \\
\xi &=  \Delta^{-1} \left( \mathbf{u} \cdot \nabla \eta - u_3 \right)
\end{align}
into \eref{xi_fo} and \eref{eta_fo}. Hence,
\begin{align} \label{variationalNu}
 \text{Nu}\{ \mathbf{u} \} - 1 = \max_{\xi} \min_{\eta} \mathcal{S} = \min_{\eta} \max_{\xi} \mathcal{S}
\end{align}
for each \textit{fixed} velocity field $\mathbf{u}$.

The formula \eref{variationalNu}, which first appeared in \cite{tobasco2017} and was further analyzed in \cite{tobasco2018}, is an exact variational characterization of the Nusselt number. It allows the optimal wall-to-wall transport problem to be stated succinctly as
\begin{align}
\label{ordering}
\max_{\mathbf{u}} Nu\{\mathbf{u}\} - 1 = \max_{\mathbf{u} }\max_{\xi} \min_{\eta} \mathcal{S} = \max_{\mathbf{u} } \min_{\eta} \max_{\xi} \mathcal{S} 
\end{align}
where $\mathbf{u}$ satisfies the given boundary conditions and intensity constraints. In particular, we note that gradient ascent may be applied with impunity to the constrained functional $\mathcal{S}_\eta$ to compute local maxima. This functional can also be used to prove {\it lower} bounds on the Nusselt number without having to solve the advection-diffusion equation.
Indeed, plugging in \textit{any} $\xi$ admissible in the previous manipulations into $\mathcal{S}_\eta$ yields the lower bound 
\begin{align}
\mathcal{S}_\eta \{\mathbf{u},\xi\} \leq Nu\{\mathbf{u}\} - 1. \label{lowerbd_test}
\end{align}
In fact, reinterpreting for the moment angle brackets as space and time averages, we note that the bound \eref{lowerbd_test} holds for time-independent flows as well, an observation that was exploited in \cite{tobasco2017} and \cite{tobasco2018} with ``branching" trial functions to prove the scaling $\max\,Nu \sim Pe^{2/3}$ up to logarithmic corrections as $Pe\to \infty$. We return to discuss this asymptotic result in the context of our numerical results much further below. Next, we consider the relationship between the wall-to-wall problem and the background method.

\subsection{Finding the background method}
\label{background_method}
%It is tempting to wonder the extent to which ordering of the infinum and suprema in \eref{ordering} matters for the computation of saddle points of $\mathcal{S}$.
%This idea was first explored by \cite{tobasco2017} but we add now to the discussion by arguing that the ordering does indeed lead to fundamentally different optimization problems.

Consider now the background method which guarantees the absolute upper bound
\begin{align}
\label{background}
\text{Nu} - 1 \leq \min_{\eta(\mathbf{x}),\mu}
\begin{cases}
\langle |\nabla \eta |^2\rangle + \mu \text{Pe}^2  &\text{if } \langle \mathcal{Q}[\mathbf{u},\xi;\eta ] \rangle  \leq  0\ \  \forall\,\mathbf{u},\xi \\
\infty & \text{otherwise} 
\end{cases}
\end{align}
where 
\begin{align}
\label{spec_const}
\mathcal{Q}[\mathbf{u},\xi;\eta ]  =   2\xi \mathbf{u}\cdot\left(  \hat{z} - \nabla \eta  \right)- |\nabla \xi |^2 - \mu | \nabla \mathbf{u}|^2 .
\end{align}
The reader may not immediately recognise this as the familiar background method as it has been applied to Rayleigh-B\'enard convection \cite{DC1996}. Nevertheless, \eref{background} does follow from applying the usual argument to the wall-to-wall problem. (The resulting bounds carry over to the time-dependent case.)
 
Let us recall the argument now. Starting with the advection-diffusion equation 
\begin{align}
\mathbf{u} \cdot \nabla T = \Delta T
\end{align}
and decomposing $T$ as $T = \xi + 1-z + \eta$ yields
\begin{align}
\label{advec_diff_sym}
 \mathbf{u} \cdot \nabla \xi + \mathbf{u} \cdot \nabla \eta = \Delta \xi  + \Delta \eta + u_3.
\end{align}
Multiplying through by $\xi$ and integrating by parts yields the balance relation
\begin{align}
\label{balance}
\langle \xi \mathbf{u} \cdot \nabla \eta + | \nabla \xi |^2 - u_3 \xi + \nabla \eta \cdot \nabla \xi\rangle = 0.
\end{align}
Now, utilising 
\begin{align}
\label{nus}
\langle w T \rangle  = \langle |\nabla T |^2 \rangle -1 =  \langle |\nabla \xi |^2 + | \nabla \eta |^2  + 2 \nabla \xi \cdot \nabla \eta \rangle,
\end{align}
we subtract twice \eref{balance}  from  \eref{nus} to conclude that
\begin{align}
\label{cfundm}
\text{Nu} - 1 = \left\langle |\nabla \eta |^2 + 2\xi \mathbf{u}\cdot\left(  \hat{z} - \nabla \eta  \right)- |\nabla \xi |^2 \right\rangle.
\end{align}
Introducing a Lagrange multiplier $\mu/2$ for the enstrophy constraint $\langle | \nabla \mathbf{u} |^2 \rangle = Pe^2$ yields
\begin{align}
\text{Nu} - 1 & = \left\langle |\nabla \eta |^2 + 2\xi \mathbf{u}\cdot\left(  \hat{z} - \nabla \eta  \right)- |\nabla \xi |^2 \right\rangle + \mu (Pe^2 - \langle | \nabla \mathbf{u} |^2 \rangle) \label{Nu_identity}\\
 & = \langle |\nabla \eta|^2 \rangle  + \mu Pe^2 + \langle \mathcal{Q}[\mathbf{u},\xi;\eta ] \rangle
\end{align}
and upon performing the operations $\min_\eta \max_{\mathbf{u},\xi}$ we deduce \eref{background}.

\subsection{A possible duality gap}
\label{poly_examp}
We can now discuss the relationship between the background method and the wall-to-wall optimal transport problem. Combining the definition of $\mathcal{S}$ from \eref{cfund} and the identity \eref{Nu_identity} we see that the background method bound \eref{background} can be alternatively written as
\begin{align}
\max_{\mathbf{u}} Nu -1 \leq \min_\eta \max_{\mathbf{u},\xi} \mathcal{S}.
\end{align}
On the lefthand side appears the wall-to-wall optimal transport problem, while on the righthand side appears the background method. Note this inequality is consistent with the results of \sref{w2w_chara} since in any case 
\begin{align}
\max_{\mathbf{u},\xi} \min_\eta \mathcal{S} \leq \min_\eta \max_{\mathbf{u},\xi} \mathcal{S}
\end{align}
regardless of the definition of $\mathcal{S}$. Now if $\mathcal{S}$ were convex in $\eta$ and jointly concave in $(\mathbf{u},\xi)$ one would be lead on general grounds via convex duality to conjecture that equality should hold between the lefthand and righthand sides above, in which case the wall-to-wall problem and the background method would turn out to be equivalent. We instead propose that the opposite situation is true and that
\begin{align}\label{inequality}
\max_{\mathbf{u}} Nu -1  \neq \min_\eta \max_{\mathbf{u},\xi} \mathcal{S} .
\end{align}
 %We caution that an analysis based solely on Euler-Lagrange equations will be \textit{insensitive} to this issue, since it requires to distinguish between critical points.
Were such an equality true, it would not preclude the possibility that these quantities achieve the same asymptotic scaling as $\text{Pe} \to \infty$, a situation suggested for 3D wall-to-wall optimal transport by the recent numerical scaling $\max\, \text{Nu} \sim \text{Pe}^{2/3}$ reported for a finite range of $\text{Pe}$ in \cite{osakaheat2}. It would also be consistent with the dimension-independent logarithmic lower bound $\max\, \text{Nu} \geq C' Pe^{2 / 3}/(\log  \text{Pe} )^{4/3}$ proved for all large enough $\text{Pe}$ in \cite{tobasco2017} and \cite{tobasco2018}.
%The relation between the background method and the wall-to-wall problem was noted previously by \cite{SouzaThesis2016} and \cite{tobasco2017}.

Let us illustrate the possibility that \eref{inequality} holds by considering how the previous manipulations operate on the polynomial
\begin{align}
\label{simple}
p(\tau_1,\tau_2,\tau_3, u , v) &= (\tau_1)^2 +  (\tau_2)^2 +  (\tau_3)^2 + q(u,v,\tau_1,\tau_2,\tau_3) ,\\
\label{simp_spec}
q &= \begin{bmatrix}
u & v
\end{bmatrix}
\begin{bmatrix}
2(1-\tau_1 +\tau_2) & \tau_3 \\
\tau_3 & 2(1-\tau_1 - \tau_2)
\end{bmatrix}
\begin{bmatrix}
u \\
v
\end{bmatrix},
\end{align}
which we see as analogous to \eref{cfund}.
The variable $\tau_1$ is analogous to the zero'th Fourier mode of $\eta$, while $\tau_2$ and $\tau_3$ are to the non-zero Fourier modes.\footnote{This polynomial was not derived as a modal truncation of $\mathcal{S}$. That would produce a more complicated example.}
The variables $u$ and $v$ are analogous to $\xi$ and $\mathbf{u}$.
The fact is that 
\begin{align}
\max_{u,v} \min_{\tau_1,\tau_2,\tau_3} p = 4/5  <  \min_{\tau_1,\tau_2,\tau_3} \max_{u,v} p = 1
\end{align}
and that the optimizers for the ``background method'' $\min$ $\max$ problem are \textit{not} saddle points of $p$. In particular, the critical point equations
\begin{align}
\label{simple_saddle}
\tau_1 &= u^2 + v^2 , \tau_2 = v^2- u^2, \text{ and } \tau_3 = -uv \\
\begin{bmatrix}
0 \\
0
\end{bmatrix}
&= 
\begin{bmatrix}
2(1-\tau_1 +\tau_2) & \tau_3 \\
\tau_3 & 2(1-\tau_1 - \tau_2)
\end{bmatrix}
\begin{bmatrix}
u \\
v
\end{bmatrix}
\end{align}
fail to be satisfied by solutions of the $\min$ $\max$ problem.

To see why this is the case, consider the background method procedure wherein the maximum occurs first. If any of the eigenvalues of \pref{simp_spec} are positive then the maximum over $u$ and $v$ yields infinity; thus we must calculate the eigenvalues of \pref{simp_spec} to see when this occurs. For fixed $(\tau_1,\tau_2,\tau_3)$ the eigenvalues of the matrix in \pref{simp_spec} are $\lambda = 2(1-\tau_1) \pm \sqrt{4 (\tau_2)^2 + (\tau_3)^2}$.  The only way these eigenvalues are nonpositive is if $2(1-\tau_1)   + \sqrt{4 (\tau_2)^2 + (\tau_3)^2} \leq 0$, or equivalently $2 \tau_1 \geq 2 +\sqrt{4 (\tau_2)^2 + (\tau_3)^2}$. Hence, 
\begin{align}
\max_{u,v}p(\tau_1,\tau_2,\tau_3, u , v)
=
 \begin{cases}
 (\tau_1)^2 + (\tau_2)^2 + (\tau_3)^2 & \text{ if } 2 \tau_1 \geq 2 +\sqrt{4 (\tau_2)^2 + (\tau_3)^2}\\
 \infty & \text{ otherwise}
 \end{cases}.
\end{align}
%We see there are no restrictions on the infinimum over $\tau_2$ and $\tau_3$ so they can be made zero, but $\tau_1 \geq 1 +\sqrt{(\tau_2)^2 + (\tau_3/2)^2}$ so its smallest value is $\tau_1 = 1$.
It follows immediately that $\min\max p = 1$, and that the minimizer satisfies $\tau_1 = 1$ and $\tau_2 = \tau_3 =0$. 
However, such $\tau_1$, $\tau_2$ and $\tau_3$ cannot be a saddle point of $p$: if $\tau_2 = \tau_3 = 0$ then from \eref{simple_saddle} we see that $u^2 = v^2$ and $uv =0$ so that $u=v=0$, but then the $\tau_1$ equation cannot be satisfied since $\tau_1 = 1 \neq  u^2 + v^2$.

Proceeding in the reverse order we find that
\begin{align}
\min_{\tau_1,\tau_2,\tau_3}p &= -(u^2 +v^2)^2 - (u^2 - v^2)^2  - (u v)^2 + 2 (u^2 + v^2).
\end{align}
The minimizing $\tau$ satisfy
\begin{align}
\tau_1 = u^2 + v^2 , \tau_2 = v^2- u^2, \text{ and } \tau_3 = -uv.
\end{align} 
The maximum over $u,v$ is given by $u = \pm \sqrt{\frac{2}{5}}$ and $v = \pm \sqrt{\frac{2}{5}}$. Hence, $ \max \min p = 4/5$. 

While in this example the background method procedure (maximum followed by minimum) yields results that are incompatible with the saddle points of \eref{simple}, the wall-to-wall procedure (minimum followed by maximum) does produce saddle points. Returning to the actual wall-to-wall optimal transport problem, we note that the optimal flow fields reported in \sref{results} exhibit non-trivial non-zero Fourier modes for the variable $\eta$, whereas in the background method optimizers must satisfy $\eta = \eta(z)$. Indeed, if $\eta$ satisfies the spectral stability constraint $\langle \mathcal{Q} \rangle \leq 0$ then so does its periodic average $\overline{\eta }$ in $x$, while by Jensen's inequality
$\langle | \frac{d}{dz} \overline{\eta} |^2 \rangle \leq \langle |\nabla \eta |^2 \rangle$ with equality if and only if $\eta = \overline{ \eta }$.
These observations strongly suggest that the conjectured gap \eref{inequality} between the wall-to-wall problem and the background method should hold, and in particular that the spectral stability constraint should fail to be satisfied by the true saddle points of $\mathcal{S}$.

\subsection{Comparison with the Howard-Busse-Malkus problem}
\label{hbm_prob}
We would be remiss if we did not additionally state the connection of the previous discussions on the wall-to-wall and background method approach with the classic Howard-Busse-Malkus approach put forth in \cite{Howard63}. To see the connection between these, start with $\mathcal{S}$ and restrict attention to incompressible flows $\nabla \cdot \mathbf{u} = 0$ with enstrophy $\langle | \nabla \mathbf{u} |^2 \rangle = \text{Pe}^2$ and functions $\eta(z)$. At this point it is useful to introduce notation for the horizontal average of a function,
\begin{align}
\overline{f} \equiv \frac{1}{\Gamma}\int_0^{\Gamma} f(x,z) dx.
\end{align}
Computing the minimum of $\mathcal{S}$ with respect to $\eta(z)$  yields the optimality condition
\begin{align}
\frac{d^2}{dz^2 } \eta  = \overline{\mathbf{u} \cdot \nabla \xi} =  \frac{d}{dz}\overline{ u_3 \xi}
\end{align}
whose solution is
\begin{align}
\eta(z) &= \int_0^z \overline{ u_3 \xi} (z') dz' - z \int_0^1\overline{ u_3 \xi} (z') dz' = \int_0^z \overline{ u_3 \xi} (z') dz' - z \langle u_3 \xi \rangle  \\
\frac{d}{dz}\eta &= \overline{ u_3 \xi}(z) - \langle u_3 \xi \rangle.
\end{align}
Employing this relation in $\mathcal{S}$ yields
\begin{align}
\label{additive}
\min_{\eta(z)} \mathcal{S} &= \left\langle 2 u_3 \xi - \left( \overline{u_3 \xi} - \langle u_3 \xi \rangle \right)^2 - |\nabla \xi |^2 \right\rangle.
\end{align}
Making the change of variables $\xi = \alpha \sigma$ for a soon to be determined scalar $\alpha$ transforms this to
\begin{align}
\label{additive_1}
\min_{\eta(z)} \mathcal{S} &= \alpha \left\langle 2 u_3 \sigma \right\rangle  -\alpha^2 \left\langle \left( \overline{u_3 \sigma} + \langle u_3 \sigma \rangle \right)^2 + |\nabla \sigma |^2 \right\rangle,
\end{align}
a quadratic function of $\alpha$.
Maximizing with respect to $\alpha$ determines the optimal choice
\begin{align}
\alpha^* =  \frac{\langle u_3 \sigma \rangle }{\left\langle \left( \overline{u_3 \sigma} - \langle u_3 \sigma \rangle \right)^2 + |\nabla \sigma |^2 \right\rangle} .
\end{align}
utilising $\alpha^*$ in the above and taking $\mathbf{u} = \text{Pe} \frac{\mathbf{v}}{\sqrt{\langle |\nabla \mathbf{v} |^2\rangle}}$ results in the multiplicative form of the functional 
\begin{align}
\label{multiplicative}
\mathcal{M}[\sigma, \mathbf{v}, \text{Pe}] &= \frac{\left( \langle v_3 \sigma \rangle \right)^2 }{\text{Pe}^{-2} \langle |\nabla \sigma |^2 \rangle \langle |\nabla \mathbf{v} |^2 \rangle + \langle \left( \overline{v_3 \sigma} - \langle v_3 \sigma \rangle \right)^2 \rangle },
\end{align}
exactly as in \cite{Howard63} under the appropriate retranscriptions.
The functional in \eref{multiplicative} is homogeneous with respect to $\sigma$ and $\mathbf{v}$, i.e., $\mathcal{M}[\lambda \sigma, \lambda \mathbf{v}, \text{Pe}]  = \mathcal{M}[\sigma, \mathbf{v}, \text{Pe}]$. It has in the past served as a starting point for the analysis of maximal heat transport, in particular leading in \cite{Howard63} to bounds on transport under the author's assumptions of homogeneity and statistical similarity. The connection of this functional to the background method has been pointed out before \cite{Kerswell98}.

%Note we do not use this multiplicative form of $\inf_{\eta(x,z)} \mathcal{S}$ in our numerical computations. 
However, we point out now that the resulting bounds on transport can be improved beyond those obtained using  \eref{multiplicative} due to the variational representation \eref{variationalNu} of the wall-to-wall problem. After a similar series of manipulations utilising all possible $\eta(x,z)$ instead of functions of $z$ alone, we deduce the improved formula 
\begin{align}
\text{Nu} - 1 = \max_{\sigma} \tilde{\mathcal{M} }[\sigma, \mathbf{v}; \text{Pe}]
\end{align}
where
\begin{align}
\label{multiplicative2}
\tilde{\mathcal{M}} [ \sigma, \mathbf{v}; \text{Pe}] &= \frac{\left( \langle v_3 \sigma \rangle \right)^2 }{\text{Pe}^{-2} \langle |\nabla \sigma |^2 \rangle \langle |\nabla \mathbf{v} |^2 \rangle + \langle \left| \nabla \Delta^{-1} \left( \mathbf{v} \cdot \nabla \sigma \right) \right|^2 \rangle }.
\end{align}
The crucial difference between \eref{multiplicative2} and \eref{multiplicative} lies in the fact that
\begin{align}
\langle \left| \nabla \Delta^{-1} \left( \mathbf{v} \cdot \nabla \sigma \right) \right|^2 \rangle  = \langle \left( \overline{v_3 \sigma} - \langle v_3 \sigma \rangle \right)^2 \rangle  + \text{strictly positive terms}
\end{align}
since $ \langle \left( \overline{v_3 \sigma} - \langle v_3 \sigma \rangle \right)^2 \rangle$ is just the zero'th Fourier mode contribution to $\langle \left| \nabla \Delta^{-1} \left( \mathbf{v} \cdot \nabla \sigma \right) \right|^2 \rangle $. Hence,
\begin{align}
\tilde{\mathcal{M}} [ \sigma, \mathbf{v}; \text{Pe}] \leq \mathcal{M} [ \sigma, \mathbf{v}; \text{Pe}]
\end{align}
for all $(\sigma , \mathbf{v})$ and the inequality is strict in most cases. Bounds on transport obtained using the functional $\tilde{\mathcal{M}}$ are, therefore, at least as tight as those that have been obtained using Howard's functional $\mathcal{M}$. We take this as additional evidence of the conjectured duality gap \eref{inequality} between the wall-to-wall problem and the background method/Howard-Busse-Malkus approach.

\section{Gradient flow}
\label{grad_flow}
The theoretical developments of the previous sections give insight into numerical methods for computing the saddle points of the functionals $\mathcal{F}$ in \eref{ffund} and $\mathcal{S}$ in \eref{cfund}.
In this section we exploit their structure to derive time-stepping methods for solving the Euler-Lagrange equations.
An advantage of the approach adopted here is that it is only necessary to have a Poisson or Stoke's solver to compute candidate optimizers to the wall-to-wall problem. 

The Euler-Lagrange equations for the wall-to-wall problem are of the form
\begin{align}
\label{abstractEL}
\mathbf{0} = \mathbf{f}(\mathbf{x}) .
\end{align}
To solve this numerically, we introduce a time-derivative on the lefthand side,  % $\dot{\mathbf{x}} = \mathbf{f}(\mathbf{x})$,
\begin{align}
\label{abstractEL2}
\dot{\mathbf{x}} = \mathbf{f}(\mathbf{x}) 
\end{align}
transforming \eref{abstractEL} into a dynamical system where every local maximum is an attracting fixed point. When applied to the functional $\mathcal{S}_{\eta} = \min_{\eta} \mathcal{S}$ from \eref{inf_eta_func} this yields its local maximizers, thereby producing saddle points for $\mathcal{S}$.

Composing \eref{abstractEL} with a locally invertible function such that $P(\mathbf{0}) = \mathbf{0}$ and then introducing a time derivative defines an alternate system $\dot{\mathbf{x}}  = P(\mathbf{f}(\mathbf{x}))$.
% \begin{align}
% \label{prec_gd}
% \dot{\mathbf{x}}  = P(\mathbf{f}(\mathbf{x})) .
% \end{align}
The danger and boon of choosing such a preconditioner is that the stability of a fixed point may change: we could be computing local maxima, minima, or saddles of our original function $\mathbf{f}$ in \eref{abstractEL}.
For example, Newton's method may be viewed as choosing the negative inverse Jabobian $J^{-1}(\mathbf{x} )$ of $\mathbf{f}$, $P(\mathbf{f}(\mathbf{x})) = - J^{-1}(\mathbf{x}) \mathbf{f}(\mathbf{x})$, along with a choice of optimal step size ($\Delta t = 1$) upon temporal discretisation.
With regards to the functional $\mathcal{F}$ we will take the preconditioner approach implemented in a way that is similar to the algorithm in \cite{Baole2015}. With regards to the Euler-Lagrange equations of \eref{fund} and \eref{cfund} we ultimately solve
\begin{align}
\label{euler_lagrange1}
0 &= \frac{ \delta \mathcal{F}}{\delta \varphi } = \Delta \theta - \mathbf{u} \cdot \nabla \theta  + u_3 \\
0 &= \frac{ \delta \mathcal{F}}{\delta \theta} = \Delta \varphi + \mathbf{u} \cdot \nabla \varphi  + u_3 \\
0 &= \frac{ \delta \mathcal{F}}{\delta \mathbf{u} } =2 \mu \Delta \mathbf{u}  - \varphi \nabla \theta + (\theta + \varphi)\hat{z} - \nabla p \\
0 &= \frac{ \delta \mathcal{F}}{\delta p  } = \nabla \cdot \mathbf{u} \\
0 &= \frac{ \delta \mathcal{F}}{\delta \mu   } = \langle  \text{Pe}^2 - | \nabla \mathbf{u} |^2 \rangle  
\end{align}
or equivalently 
\begin{align}
\label{euler_lagrange2}
0 &= \frac{1}{2} \frac{ \delta \mathcal{S}}{\delta \xi} = \Delta \xi - \mathbf{u} \cdot \nabla \eta  + u_3 \\
0 &= \frac{1}{2} \frac{ \delta \mathcal{S}}{\delta \eta } = -\Delta \eta + \mathbf{u} \cdot \nabla \xi  \\
0 &= \frac{1}{2} \frac{ \delta \mathcal{S}}{\delta \mathbf{u} } = \mu \Delta \mathbf{u}  - \xi \nabla \eta + \xi \hat{z} + \frac{1}{2} \nabla p \\
0 &= \frac{ \delta \mathcal{S}}{\delta p  } = \nabla \cdot \mathbf{u} \\
0 &= \frac{ \delta \mathcal{S}}{\delta \mu   } = \langle  \text{Pe}^2 - | \nabla \mathbf{u} |^2 \rangle  .
\end{align}

\subsection{Gradient ascent methods}
\label{ascent_methods}
Here we outline various methods for solving the Euler-Lagrange equations of the wall-to-wall problem described above. One method is to evolve
\begin{align}
0 = \frac{ \delta \mathcal{F}}{\delta \varphi } , \quad
0 = \frac{ \delta \mathcal{F}}{\delta \theta} , \quad
\partial_\tau \mathbf{u} = \frac{ \delta \mathcal{F}}{\delta \mathbf{u} } , \quad
0 = \frac{ \delta \mathcal{F}}{\delta p } , \text{\ \ and \ \ }
\partial_\tau \mu = \frac{ \delta \mathcal{F}}{\delta \mu  }
\end{align}
forward in time.
This strictly enforces the advection-diffusion equation and its adjoint at every time-step, Utilising $\frac{ \delta \mathcal{F}}{\delta \mathbf{u} }$ to compute corrections to the flow field.
This approach was taken in \cite{osakaheat}, and the rate limiting step is the solution of the advection-diffusion equation and its adjoint.
In the present work, we take a different approach and compute numerical solutions to \eref{euler_lagrange1} and \eref{euler_lagrange2} Utilising two different algorithms.

The first of our algorithms involves a time-stepping procedure of the form
\begin{align}
\label{alg_1}
\partial_\tau \theta &= \frac{ \delta \mathcal{F}}{\delta \varphi } , \quad
\partial_\tau \varphi = \frac{ \delta \mathcal{F}}{\delta \theta} , \quad
\partial_\tau \mathbf{u} = \frac{ \delta \mathcal{F}}{\delta \mathbf{u} } , \quad
0 = \frac{ \delta \mathcal{F}}{\delta p } , \text{\ \ and \ \ }
\partial_\tau \mu = \frac{ \delta \mathcal{F}}{\delta \mu  }.
\end{align}
This procedure may be understood as follows.
For fixed $\mathbf{u}$, the equations for $\partial_\tau \theta$ and $\partial_\tau \varphi$ evolve towards the steady state solutions $\frac{ \delta \mathcal{F}}{\delta \varphi } = 0$ and $\frac{ \delta \mathcal{F}}{\delta \theta } = 0$, hence these evolutions are a relaxation of fully solving the Euler-Lagrange equations for $\theta$ and $\varphi$.
The equation for $\mu$ guarantees that we flow towards a flow field with the desired enstrophy $\text{Pe}$.
The last condition (for a fixed $\theta$ and $\varphi$) evolves to a solution of the optimality condition.
We enforce incompressibility at every time-step and evolve all fields at once. %hoping to converge.

There is an alternative description of this algorithm in terms of $\mathcal{S}$.
Focusing on the $\theta$ and $\varphi$ equations, we see that evolving
\begin{align}
\partial_\tau \theta &= \frac{ \delta \mathcal{F}}{\delta \varphi }  \text{ \ \ and \ \ }
\partial_\tau \varphi = \frac{ \delta \mathcal{F}}{\delta \theta}
\end{align}
is equivalent to evolving 
\begin{align}
\partial_\tau \xi &= \frac{ \delta \mathcal{S}}{\delta \xi}  \text{ and }
-\partial_\tau \eta = \frac{ \delta \mathcal{S}}{\delta \eta}
\end{align}
after taking sums and differences (making use of $\theta = \xi + \eta$ and $\varphi  = \xi - \eta$) and rescaling time.
Thus our time-stepping procedure for $\theta $ and $\varphi$ is equivalent to simultaneously applying gradient ascent for the concave variable $\xi$ and gradient descent for the convex variable $\eta $ in the $\mathcal{S}$ functional.
This is similar to the philosophy of \cite{Baole2015}.
In fact, the only modification required to compute two-dimensional no-slip background fields would be to project the $\eta$ variable to the zero'th Fourier mode at each time-step, by taking the $x$ average of the righthand side of the $\eta$ equation.

The second time-stepping method involves computing the saddles of $\mathcal{S}$ via
\begin{align}
\label{alg_2}
\partial_\tau \xi &= \frac{ \delta \mathcal{F}}{\delta \xi } , \quad 
0 = \frac{ \delta \mathcal{F}}{\delta \eta} , \quad
\partial_\tau \mathbf{u} = \frac{ \delta \mathcal{F}}{\delta \mathbf{u} } , \text{ \ \ and \ \ }
0 = \frac{ \delta \mathcal{F}}{\delta p } 
\end{align}
while fixing $\mu$. The resulting enstrophy depends implicitly on $\mu$.
%There is no need to evolve $\mu$ since a fixed $\mu$ implicitly enforces the enstrophy constraint.
Enforcing $\frac{ \delta \mathcal{F}}{\delta \eta} = 0$  at each time-step yields gradient ascent for the local optima of \eref{inf_eta_func}, an unconstrained variational problem.
The reason why this algorithm is efficient is due to the many existing algorithms for quick inversion of the Laplacian and the Helmholtz operator.  

We found numerically that both \eref{alg_1} and \eref{alg_2} yield the same results.
We also implemented various other ascent procedures with different preconditioners.
For example, we evolved equations of the form
\begin{align}
\label{alg_3}
-\Delta \partial_\tau \theta &= \frac{ \delta \mathcal{F}}{\delta \varphi } ,  \quad
-\Delta \partial_\tau \varphi = \frac{ \delta \mathcal{F}}{\delta \theta} , \quad
\partial_\tau \mathbf{u} = \frac{ \delta \mathcal{F}}{\delta \mathbf{u} } , \quad
0 = \frac{ \delta \mathcal{F}}{\delta p } , \text{ \ \ and \ \ }
\partial_\tau \mu = \frac{ \delta \mathcal{F}}{\delta \mu  }
\end{align}
forward in time, as well as
\begin{align}
\label{alg_4}
-\Delta \partial_\tau \theta &= \frac{ \delta \mathcal{F}}{\delta \varphi } ,  \quad
-\Delta \partial_\tau \varphi = \frac{ \delta \mathcal{F}}{\delta \theta} , \quad
- \Delta \partial_\tau \mathbf{u} = \frac{ \delta \mathcal{F}}{\delta \mathbf{u} } , \quad
0 = \frac{ \delta \mathcal{F}}{\delta p } , \text{ \ \ and \ \ }
\partial_\tau \mu = \frac{ \delta \mathcal{F}}{\delta \mu  },
\end{align}
evolving different components on different time-scales.
Amongst all of our results, the ones presented in \sref{results} maximize $\text{Nu}$ for a given $\text{Pe}$.

\subsection{Temporal discretisation}
\label{time_evo}
Each of the gradient ascent procedures described in \sref{grad_flow} are of the form
\begin{align}
\dot{x} &= \mathcal{L}x + \mathcal{N}(x) + f,
\end{align}
where $x$ is the state vector, $\mathcal{L}$ is an ``easily" invertible linear operator, $\mathcal{N}$ is a nonlinear operator, and $f$ is a forcing function.
%Both the advection-diffusion/adjoint equation and the gradient-ascent procedures generate these evolutions.
We follow \cite{Viswanath2013414} and consider linear multi-step schemes as follows:
\begin{align}
\frac{1}{\Delta \tau}\left(\gamma x^{n+1} + \sum_{j=0}^{s-1} a_j x^{n-j} \right) &= \sum_{j=0}^{s-1}b_j \mathcal{N}(x^{n-j}) + \mathcal{L} x^{n+1} + f^{n+1}.
\end{align}
where $s$ is the order of the time-stepping scheme,  $a_i$ and $b_i$ are parameters, and $\Delta t$ is the time-step size. The parameters values for orders $s=1,2$ and $3$ are 
\begin{align}
s &= 1, \hspace{4mm} \gamma = 1,  \hspace{4mm} a_0 = - 1, \hspace{4mm} b_0 = 1 \\
s &= 2,\hspace{4mm}  \gamma = 3/2,\hspace{4mm} a_0 = - 2, \hspace{4mm} a_1 = 1/2, \hspace{4mm} b_0 = 2, \hspace{4mm} b_1 = -1 \\
s &= 3,\hspace{4mm}  \gamma = 11/6, \hspace{4mm} a_0 = - 3, \hspace{4mm} a_1 = 3/2, \hspace{4mm} a_2 = -1/3\hspace{4mm} b_0 = 3, \hspace{4mm} b_1 = -3, \hspace{4mm} b_2 = 1.
\end{align}

For example with $s=1$ and the advection-diffusion equation,
\begin{align}
 \partial_\tau \theta &= -\mathbf{u} \cdot \nabla \theta + \Delta \theta + u_3 
\end{align}
we use
 \begin{align}
\left( \Delta - \frac{1}{\Delta \tau} \mathbf{I} \right) \theta^{n+1} &= \mathbf{u} \hspace{1mm}^{n} \cdot \nabla \theta^{n} - w^{n} - \frac{1}{\Delta \tau}\theta^n .
 \end{align}
Thus for each time-step we must solve a modified Poisson's equation of the form
 \begin{align}
  \label{mpoisson}
 \left( \Delta - c \mathbf{I} \right) \theta &= f
 \end{align}
 where $c\geq 0$ and we have made the transcriptions
 \begin{align}
 \theta^{n+1} \mapsto \theta \text{ , }
 \frac{1}{\Delta \tau}  \mapsto c \text{ , and } 
  \mathbf{u}\hspace{1mm}^{n} \cdot \nabla \theta^{n} - w^{n} - \frac{1}{\Delta t}\theta^n  \mapsto f.
 \end{align}
 Analogous discretisations are used for $\varphi, \eta ,$ and $\xi$. 
For updating the optimality condition with $s=1$ one option is to use 
\begin{align}
\left( \mu \Delta - \frac{1}{\Delta \tau} \mathbf{I} \right) \mathbf{u}\hspace{1mm}^{n+1}  &= -\left( \varphi^{n}+\theta^{n}\right)\hat{e}_3 + \varphi^n \nabla \theta^n - \frac{1}{\Delta \tau}\mathbf{u}\hspace{1mm}^{n}  + \nabla p^{n+1} \\
\nabla \cdot \mathbf{u}\hspace{1mm}^{n+1}  &= 0.
 \end{align}
Each time-step involves solving a modified Stokes equation
 \begin{align}
 \label{mstokes}
  \left( \Delta - c \mathbf{I} \right) \mathbf{u} &= \mathbf{f} + \nabla p \\
 \nabla \cdot \mathbf{u} &= 0
 \end{align}
 where $c\geq 0$ and we have made the transcriptions
 \begin{align}
 \mathbf{u}\hspace{1mm}^{n+1}  \mapsto \mathbf{u} \text{ , }
 p^{n+1} \mapsto p \text{ , }
 \frac{1}{\Delta \tau}  \mapsto c \text{ , and }
 \left( \varphi^{n}+\theta^{n}\right)\hat{e}_3 + \varphi^n \nabla \theta^n - \frac{1}{\Delta \tau}\mathbf{u}\hspace{1mm}^{n}   &\mapsto \mathbf{f}.
 \end{align}
 
We solve these boundary value problems by Utilising a pseudo-spectral method, where the wall-bounded direction is represented by Chebyshev polynomials and the periodic directions are represented by Fourier Series; however our use of Chebyshev polynomials utilizes spectral integration in the same way as \cite{Viswanath2015159}.
The Stokes equation \eref{mstokes} was solved Utilising the Kleiser-Schuman algorithm \cite{ksalg}.
See \aref{disc_details} for details regarding implementation. A theoretical discussion of the approach is in \cite{Viswanath2015159}. Let us highlight some of the benefits of spectral integration here:
\begin{enumerate}
\item Memory efficiency: The total memory occupied is linear in the number of points required to describe the state variable $\mathbf{u}$.
\item Speed: Solving the linear equations involve only tridiagonal matrices which are put into LU form at the beginning of the gradient ascent procedure.
\item Discretisation Accuracy: Everything is represented using Fourier and Chebyshev modes, allowing for spectral accuracy.
\item Machine Accuracy: Utilising spectral integration in factored form (see \aref{disc_details}) allows one to avoid taking wall-bounded derivatives and only take derivatives in periodic direction, without passing to the nodal domain.
\end{enumerate}

At this point all the pieces are place to compute flow fields maximizing heat transfer. By Utilising gradient ascent or other time-marching schemes of \sref{ascent_methods} it is possible to adapt existing Rayleigh-B\'enard codes to find steady maximizing flow fields and compare their thermal transfer to that of natural flows.

\section{Two-dimensional computations}
\label{results}
For every decade of P\'eclet we computed approximately 20 logarithmically spaced solutions to the Euler-Lagrange equations \eref{euler_lagrange1} and \eref{euler_lagrange2}.
We then performed numerical differentiation of $\log{\text{Nu}}$ and $\log{\Gamma}$ with respect to $\log{\text{Pe}}$ to examine (local) scaling relations.
The largest enstrophy satisfied $\text{Pe} \approx 10^5$, with an $x$ Fourier, $z$ Chebyshev grid size of $512\times 1025$.
The time-stepping code slowed substantially at larger $\text{Pe}$.

In addition to numerical continuation from small P\'eclet ($\text{Pe} \approx 0$), we started at different points in function space in an attempt to find more optimal solutions.
We utilized solutions to other related variational problems as ``educated guesses" for new flows at which we started our gradient ascent procedure.
We computed solutions to the Euler-Lagrange equations for fixed aspect ratio $\Gamma$ as well as optimizing over all $\Gamma$.
The solutions presented here achieved the largest transport. 
%(One should, of course, keep in mind that no numerical search will ever be exhaustive and there could be other extrema that transport more heat than the ones presented here.)

\begin{figure}
\begin{center}
\includegraphics[width=0.7\textwidth]{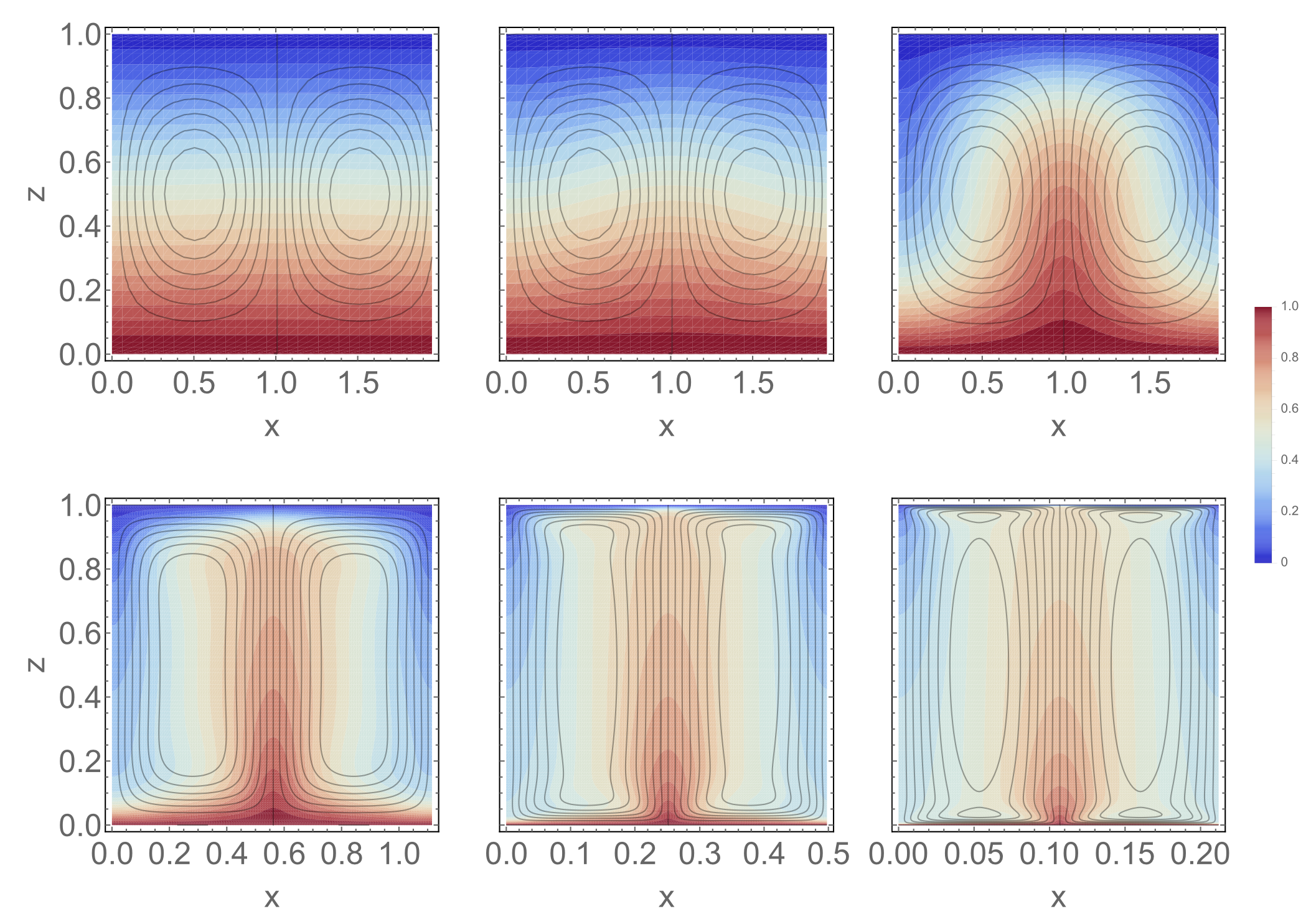}
\caption{Optimal no-slip solutions for different enstrophy budgets in a single cell. The black contour lines are the streamlines and the colours represent the temperature field. From left to right, top to bottom the P\'eclet numbers are $4.0 \times 10^{-1},  4.0 \times 10^0, 4.0 \times10^{1}, 4.0\times10^{2}, 4.0 \times 10^{3}, 4.0 \times10^{4}$. The domain size in the horizontal direction $x$ shrinks as the enstrophy budget increases.}
\label{fig:z_ns}
\end{center}
\end{figure}

\fref{fig:z_ns} shows visualizations of the temperature field and the stream function with no-slip boundary conditions.
In the low P\'eclet regime the solutions are convection cells while at higher Pe ``recirculation zones" develop on the bottom and top boundaries.
Additionally the optimal aspect ratio of a cell size shrinks with ever increasing $\text{Pe}$.
For a fixed aspect ratio optimal solutions contain a multiplicity of convection cells.
One of the benefits of our numerical approach is the automatic computation of the optimal domain size with little additional overhead. 

\begin{figure}
\begin{center}
\includegraphics[width=1.0\textwidth]{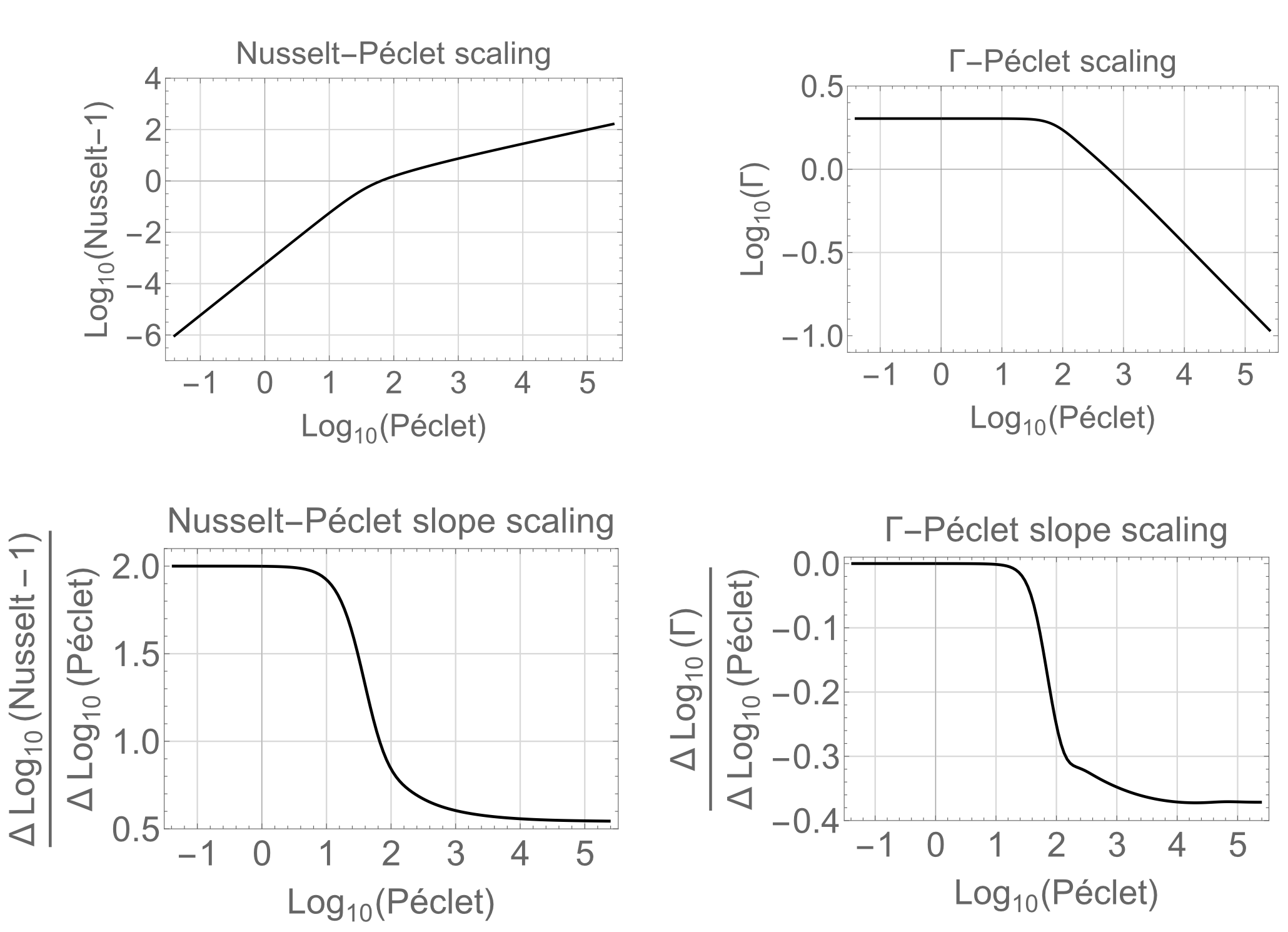}
\caption{Computed optimal Nusselt number (Nu) and aspect ratio ($\Gamma$) as a function of the enstrophy budget ($\text{Pe}$), for no-slip boundary conditions. Top Left: Log-Log plot of Pe vs Nu-1. Bottom Left: The instantaneous slope of the top left plot, $\text{Pe}$ vs $d \log(\text{Nu}-1) / (d \log \text{Pe} )$. Top Right: Log-Log plot of $\text{Pe}$ vs $\Gamma$. Bottom Right: The instantaneous slope of the top right plot, $\text{Pe}$ vs $d \log(\Gamma) / (d \log \text{Pe} )$. The last instantaneous slope for the bottom left plot is $ 0.544$ and the last instantaneous slope for the bottom right plot is $-0.371$. The largest computed $\text{Pe} = 2.5 \times 10^5$ corresponding to $\mu = 1.4 \times 10^{-9}$. }
\label{fig:ns_scalings}
\end{center}
\end{figure}

In \fref{fig:ns_scalings} we report the $\text{Nu}$-$\text{Pe}$ and $\Gamma$-$\text{Pe}$ relations and local scaling relations for the best known optimizers (for no-slip boundary conditions).
After leaving the ``linear" regime where $\text{Nu} \sim \text{Pe}^2$ we enter a ``fully nonlinear" regime where $\text{Nu} \sim \text{Pe}^{0.54}$.
In view of the results of \cite{tobasco2017} and \cite{tobasco2018} this scaling cannot persist for the globally maximal  transport at sufficiently large $\text{Pe}$.
Nevertheless, it does appear to be optimal for the computed range of P\'eclet in two spatial dimensions.\footnote{The $\text{Nu} \sim \text{Pe}^{0.54}$ optimal transport scaling in this regime was first reported in \cite{SouzaThesis2016} and thereafter confirmed by independent computation by \cite{osakaheat} and \cite{kerswell2019}.  %In view of the subsequent computations reported in \cite{osakaheat2}, however, we now know that these two-dimensional flows are not global optimizers in three spatial dimensions.
}

In the linear regime there is little change in the optimal aspect ratio, i.e., the optimal $\Gamma \approx \text{constant}$ while in the fully nonlinear regime a nontrivial $\Gamma \sim \text{Pe}^{-0.37}$ scaling emerges.
Interestingly, the local exponent exhibits a somewhat oscillatory relaxation rather than a perhaps more expected monotonic convergence.

\subsection{Singular value decomposition}

\begin{figure}
\begin{center}
\includegraphics[width=1.0\textwidth]{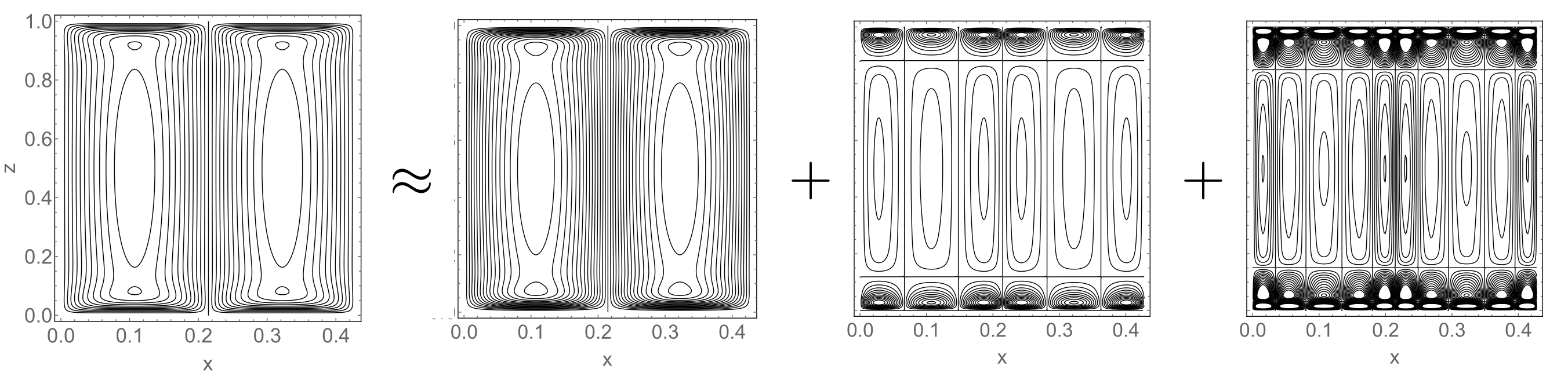}
\caption{The stream function $\psi$ and its first three modes at $\text{Pe} \approx 2.4 \times 10^4$. The modes are ordered left to right by the magnitude of their singular values starting with the largest.}
\label{fig:decomp_stream}
\end{center}
\end{figure}

Given the structure of the solutions depicted in \fref{fig:z_ns} one may wonder if the corresponding fields are to first approximation separable. With regards to the stream function $\psi$ defined by $(-\partial_z \psi, \partial_x \psi ) = (u_1 , u_3)$ this would mean
\begin{align}
\psi(x,z) \approx \Psi(z) \phi(x).
\end{align}
We investigate this possibility by minimizing the functional 
\begin{align}
A[\Psi,\phi ] &= \frac{1}{\Gamma} \int_{0}^{\Gamma}\int_{0}^1 (\psi(x,z) - \Psi(z)\phi(x) )^2  dx dz
\end{align}
subject to appropriate constraints on $\Psi(z)$ and $\phi(x)$, and similarly for other fields such as $\eta$ or $\xi$.\footnote{E.g., $\phi(x)$ should be periodic and $\Psi(z)$ should satisfy no-slip boundary conditions.}
If the minimal value is sufficiently small then we may say that the flow fields are \textit{nearly} separable.

Upon discretization,
\begin{align}
\label{approx_int}
A[\Psi,\phi ] &\approx \sum_{i,j} (\psi_{ij} - \Psi_{j} \phi_{i} )^2 w_{ij} \\
\psi_{ij} = \psi(x_i,z_j)  \text{, }
\phi_i &= \phi(x_i) \text{ , and }
\Psi_{j} = \Psi(z_j)  
\end{align}
where $x_i = \Gamma i / n$ for $i = 0,...,n-1$ and $z_j = \frac{1}{2}\left( 1+\cos(\pi j/m ) \right)$ for $j = 0,...,m$ are the collocation points of the Fourier and Chebyshev discretisations, and $w_{ij}$ are weights for approximating the integral by a discrete sum.
Spectral accuracy for the approximation \eref{approx_int} was obtained by setting
\begin{align}
w_{ij} &= \Delta x_i \Delta z_j 
\text{ with } 
\Delta x_i = \frac{1}{n} \text{ and } 
\Delta z_j = \frac{1}{2} \sin(a_j) \frac{1}{m} \sum_{\ell = 1}^{m-1} \sin(\ell a_j ) \frac{(1- \cos(\ell \pi) )}{\ell}
\end{align}
where $a_j = \pi j/m $ for $j= 0, ..., m$, see \cite{Boyd}.
Since we are dealing with the average value of the integral there is no factor of $\Gamma$ in the $\Delta x_i$.

In the discrete problem corresponding to \eref{approx_int} we minimised the weighted Frobenius norm of the matrix $\psi_{ij}$ with respect to an outer product decomposition, hence for $w_{ij} = 1$ the solution to the discrete problem is to take $\Psi$ and $\phi$ to be equal to the largest singular vectors of the matrix $\psi_{ij}$ multiplied by the largest singular value.
For $w_{ij} \neq 1$ we rescaled the problem  taking $\psi_{ij} ' = \psi_{ij} \sqrt{w_{ij}}$ and found the largest singular vectors of $\psi_{ij}'$.
The difference between uniform and nonuniform $w_{ij}$ may be interpreted as minimising with respect to different weighted energy norms and, in our numerical experiments, we found little qualitative difference between utilising one over the other.
All singular value decompositions reported here were computed using $w_{ij} = 1$.

\begin{figure}
\begin{center}
\includegraphics[width=1.0\textwidth]{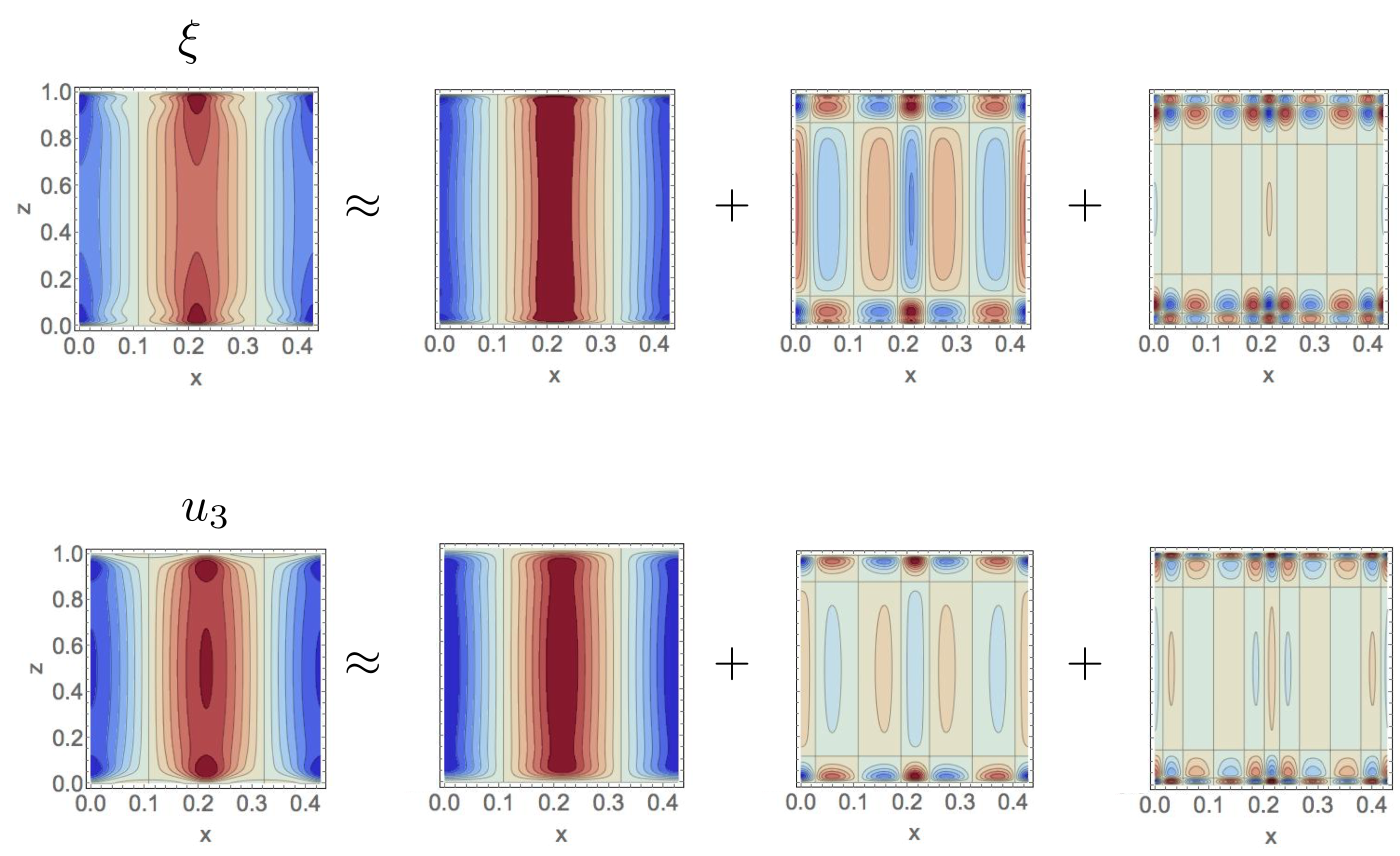}
\caption{[Colour Online] The field $\xi = \theta + \varphi$ (top), the vertical velocity $u_3$, and their first three modes at $\text{Pe} \approx 2.4 \times 10^4$. The outer products are ordered with respect to their singular values.}
\label{fig:decomp}
\end{center}
\end{figure}

\fref{fig:decomp_stream} shows the first three modes in the singular value decomposition of the stream function $\psi$. \fref{fig:decomp} shows the same for the $\xi$ field and the vertical velocity field $u_3$.  The modes with smaller singular values may be viewed as a preliminary manifestation of a branching-like pattern, perhaps similar to the ones constructed in \cite{tobasco2017} and \cite{tobasco2018}. The structures corresponding to the largest singular value are the direct analogue for the wall-to-wall problem of the single-wavenumber solutions for the Howard-Busse-Malkus problem produced in \cite{Howard63}. %While in the Howard-Busse-Malkus problem single wavenumber solutions arise via a Fourier basis, apparently in the wall-to-wall problem the  a basis other than sines and cosines.

As it turns out, these dominant structures carry $\approx 99 \%$ of the overall heat transport in the computed range of $\text{Pe}$. More precisely, writing
\begin{align}
\psi(x,z) &\approx \Psi(z) \phi_1(x), \\
\xi(x,z) &\approx \Xi(z) \phi_2(x), \\
\label{other_nus}
N_1 &= \langle (\partial_x \psi) \xi \rangle, \\
N_2 &= \langle \Psi  (\partial_x \phi_1 ) \Xi  \phi_2 \rangle,
\end{align}
our solutions achieved $(N_1 - N_2)/N_1 \leq 0.01$ uniformly over all computed P\'eclet. (This error was even less for smaller P\'eclet.) The fact that \pref{other_nus} is equal to $\text{Nu}-1$ for solutions to the wall-to-wall problem follows from the Euler-Lagrange equation \pref{euler_lagrange2}.

\begin{figure}
\begin{center}
\includegraphics[width=0.8\textwidth]{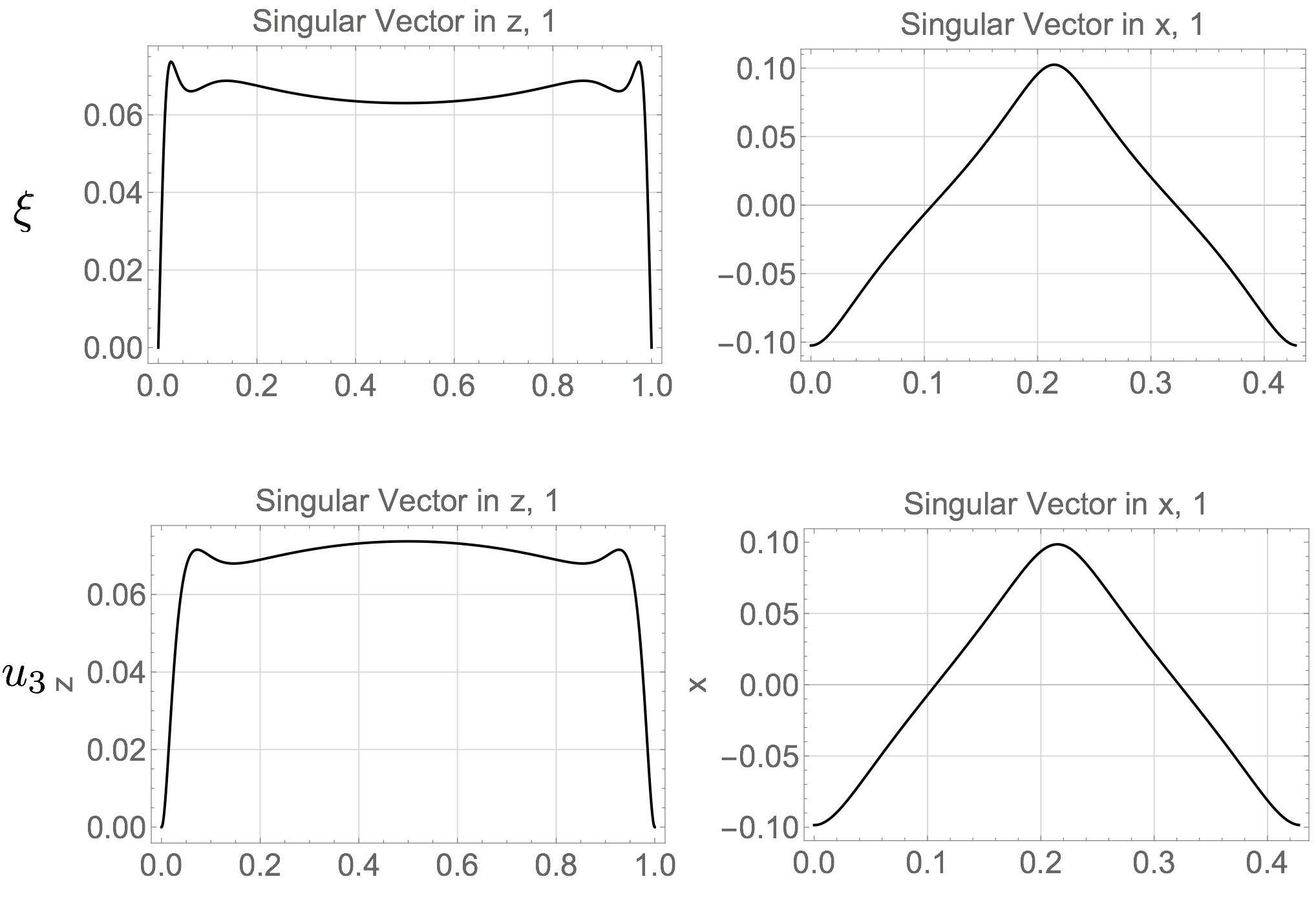}
\caption{The first singular vectors in the singular value decomposition of the $\xi$ field (top) and the vertical velocity $u_3$ (bottom). The respective products of these singular vectors yields the first approximations in \fref{fig:decomp}. The P\'eclet number $\approx 2.4\times 10^4$. }
\label{fig:sing_w_xi}
\end{center}
\end{figure}

\fref{fig:sing_w_xi} depicts the first singular vectors for $\xi$ and $u_3$. While in the Howard-Busse-Malkus problem the corresponding $x$ dependence is perfectly sinusoidal, for the wall-to-wall problem the $x$ dependence of the first singular vectors resemble Jacobi elliptic functions. The similarity between $\phi_1$ and $\phi_2$ motivates the ``separable ansatz''
 \begin{align}
u_1(x,z) &= -\Psi'(z) \phi(x) ,\\
u_3(x,z) &= \Psi(z) \phi'(x) ,\\
\xi (x,z) &= \Xi(z) \phi'(x)
\end{align}
which we consider further in \sref{tp_ans} below. There we derive upper bounds extending the single wavenumber analysis from \cite{Howard63} to the wall-to-wall problem. We will see that the functional from \sref{hbm_prob} can be bounded from above by $ \text{Pe}^{6/11} = \text{Pe}^{0. \overline{54}}$ amongst all separable ansatzes in accord with the numerically computed $\text{Nu} \sim \text{Pe}^{0.54}$ scaling.

\section{Upper bound}
\label{tensor_product_upper_bounds}
As discussed in the previous section, the flow and temperature fields found to maximize heat transport are nearly separable and achieve the scaling $Nu \sim Pe^{0.54}$ within the range of computed $\text{Pe}$. Furthermore, most of the heat transport scaling was attained via a separable approximation.
We now make use of the theoretical developments from \sref{f_a_c} to argue for the upper bound $Nu \leq C Pe^{0.\overline{54}}$ under the assumption of a perfectly separable ansatz. We do this by rederiving the Howard functional from \cite{Howard63} (as in \sref{hbm_prob}) and assuming a separable ansatz at the outset. It should be emphasized that this is \textit{not} a rigorous proof for the observed numerical scaling from the previous section since it must still be proven that all solutions in the range $\text{Pe}  \lesssim 10^5$ are indeed separable, or what is more likely the case, that their non-separable part contributes negligibly to transport. 

\subsection{Upper bounds within a separable ansatz}
\label{tp_ans}
To obtain upper bounds on the Nusselt number within a separable ansatz, we start by recalling that the Howard-Busse-Malkus problem bounds the wall-to-wall problem from above.
Indeed, as was shown in \sref{w2w_chara} and \sref{hbm_prob}, 
\begin{align}
\label{mult_relations}
\max_{\mathbf{u}} \text{Nu} -1 = \max_{\mathbf{u}, \xi} \min_{\eta} \mathcal{S} = \max_{\mathbf{v}, \sigma } \tilde{\mathcal{M}} \leq \max_{\mathbf{v}, \sigma } \mathcal{M}
\end{align}
since the minimum is only made smaller upon enlarging the class of admissible functions. Thus we are led to consider the functional 
\begin{align}
\label{multiplicative_conv}
\mathcal{M}[\sigma, \mathbf{v}, \text{Pe}] &= \frac{\left( \langle v_3 \sigma \rangle \right)^2 }{\text{Pe}^{-2} \langle |\nabla \sigma |^2 \rangle \langle |\nabla \mathbf{v} |^2 \rangle + \langle \left( \overline{v_3 \sigma} - \langle v_3 \sigma \rangle \right)^2 \rangle }
\end{align}
from \sref{hbm_prob}, while also restricting ourselves to the separable ansatz  
\begin{align}
\label{sep1}
u_1(x,z) &= -\Psi'(z) \phi(x) \\
\label{sep2}
u_3(x,z) &=\Psi(z) \phi'(x) \\
\label{sep3}
\xi(x,z) &= \Xi (z) \phi'(x)
\end{align}
where $'$ denotes an ordinary derivative. Note, while the argument that follows is similar to those of \cite{Howard63} and \cite{DC1996}, it is more general in scope since here we allow $\phi(x)$ to be \textit{any} periodic function rather than some well-chosen Fourier mode. Our choice to study this more general case is motivated by the previous discussion of numerical results, which showed how in the wall-to-wall problem non-sinusoidal $\phi(x)$ arise. Our goal now is to bound the functional $\mathcal{M}$ from above under the assumptions of separability \pref{sep1}-\pref{sep3}. Again, we note that this does not provide the unconditional upper bound on $\mathcal{M}$ required to deduce bounds on $\text{max }\text{Nu}$ according to \eref{mult_relations}. What we lack is a proof that \pref{sep1}-\pref{sep3} are indeed valid assumptions over the given range of $\text{Pe}$. Nevertheless, we proceed.

Note that the velocity field is incompressible, and normalize $\phi$ by requiring
\begin{align}
\label{norma}
\overline{(\phi')^2} = 1.
\end{align}
Given the homogeneity of the functional we may impose that
\begin{align}
\label{net_flux}
\langle u_3 \xi \rangle &= \int_0^1 \Psi(z) \Xi(z) dz = 1.
\end{align}
Given these, we seek to bound the denominator of \eref{multiplicative_conv} from below so as to produce the desired upper bound.

Note the identities
\begin{align}
\langle u_3 \xi \rangle &= \int_0^1 \Psi(z) \Xi(z) dz ,\\
\langle | \nabla \mathbf{u} |^2 \rangle &=  \int_0^1\left( \overline{\phi^2 }  ( \Psi'' )^2  + 2 ( \Psi')^2 + \overline{ ( \phi'' )^2 } (\Psi)^2 \right) dz ,\\
\langle | \nabla \xi |^2 \rangle &= \int_0^1 ( \Xi' )^2 + \overline{(\phi'')^2} \Xi^2 dz ,\\
\langle ( \overline{u_3 \xi} - \langle u_3 \xi \rangle )^2\rangle &= \int_0^1 \left( \Psi \Xi - 1 \right)^2 dz
\end{align}
where for convenience we have abbreviated $\int_0^1 f(z) dz$ as $\int f$.
Using these we obtain the lower bound
\begin{align}
\langle |\nabla \mathbf{u}|^2\rangle \langle  | \nabla \xi |^2 \rangle &\geq  \overline{\phi^2} \int ( \Psi'' )^2 \int (\Xi')^2 + \overline{ \left( \phi''\right)^2} \int \Psi^2 \int (\Xi')^2 + \overline{\phi^2} \int (\Psi'')^2  \overline{ \left( \phi''\right)^2} \int \Xi^2 \\
&\geq \overline{\phi^2} \int (\Psi'')^2 \int (\Xi')^2 + \sqrt{\overline{ \left( \phi''\right)^2} \int \Psi^2 \int (\Xi')^2  \overline{\phi^2}  \int (\Psi'')^2 \overline{ \left( \phi''\right)^2} \int \Xi^2}
\end{align}
by an elementary Young's inequality.
The interpolation inequality 
\begin{align}
\label{gn}
\left[ \overline{(\phi'')^2} \right] \left[ \overline{\phi^2} \right]\geq \left(\overline{ \phi \phi''} \right)^2=  \left( \overline{(\phi')^2} \right)^2 = 1
\end{align}
follows by combining Cauchy-Schwarz, integration by parts, and the normalization \eref{norma}.
Similarly, there holds
\begin{align}
\int \Xi^2 \int \Psi^2 \geq 1
\end{align}
by Cauchy-Schwarz and the net flux constraint. 

Proceeding, we deduce that 
\begin{align}
\langle |\nabla \mathbf{u}|^2\rangle \langle | \nabla \xi |^2\rangle &\geq  
\overline{\phi^2} \int (\Psi'')^2 \int (\Xi')^2 + \sqrt{ \overline{ \left( \phi''\right)^2} \int (\Xi')^2    \int (\Psi'')^2 } \\
&\geq 
\left( \overline{\phi^2} \int (\Psi'')^2 \int (\Xi')^2 \right)^{1/3} \left( \overline{ \left( \phi''\right)^2} \int (\Xi' )^2   \int (\Psi'')^2  \right)^{1/3} \\
&\geq \left( \int (\Psi'')^2 \int (\Xi')^2 \right)^{2/3}
\end{align}
by way of Young's inequality in the form $a + b \geq 3^{1/3} \left(\frac{2}{3}\right)^{2/3} a^{1/3} b^{2/3}$ and \eref{gn}.
Utilising Howard's lemma (the estimate referred to as such in \cite{DC1996}), we can bound the remaining term by 
\begin{align}
\int \left( \Psi \Xi - 1 \right)^2 \geq  \left( \int (\Psi'')^2 \int (\Xi')^2 \right)^{-1/4}.
\end{align}
Taking $\delta = \left( \int (\Psi'')^2 \int (\Xi')^2 \right)^{-1/4}$ we see that the denominator of the functional \eref{multiplicative_conv} --- in the separable ansatz --- is bounded below by
\begin{align}
\delta + \text{Pe}^{-2} \frac{1}{\delta^{8/3} } \gtrsim \text{Pe}^{-6/11}
\end{align}
after minimizing over $\delta$.

Hence, the functional $\mathcal{M}$ is bounded from above by $\text{Pe}^{6/11} = \text{Pe}^{0. \overline{54} }$ in the separable ansatz \pref{sep1}-\pref{sep3}, in accord with the numerically computed solution scaling of $\text{Pe}^{0.54}$. While this is suggestive of the bound $\text{Nu} \leq C \text{Pe}^{6/11}$, it remains to be shown that the separable ansatz holds up to terms negligible in their transport for the given range of $\text{Pe}$.

\section{Summary and Conclusions}
We developed and applied time-stepping methods for solving the wall-to-wall problem for steady flows.
Along the way we developed theoretical tools illuminating connections between the wall-to-wall problem and both the background method of \cite{DC1996} and Howard's classic formulation of heat transport bounds for Rayleigh-B\'enard convection from \cite{Howard63}.
We computed optimal steady two-dimensional flow fields with no-slip boundary conditions whose heat transport scaling is $\text{Nu} \sim \text{Pe}^{0.54}$ for P\'eclet numbers between $10^3$ and $10^5$.
Upon examining these two-dimensional flows we computed their
singular value decomposition, revealing concentration onto one dominant mode.
This motivated considering consequences of the additional assumption that the flow fields are separable. Within such an ansatz, we proved a conditional upper bound scaling as $\text{Pe}^{6/11}$ corresponding to the $\text{Ra}^{3/8}$ ``single wavenumber" upper bound scaling for Rayleigh-B\'enard with no-slip boundary conditions.

In light of \cite{tobasco2017} and \cite{tobasco2018}, however, these scalings for the optimal wall-to-wall problem cannot persist for global optimisers as $\text{Pe} \rightarrow \infty$.
There it was proved that wall-to-wall optimal transport in two- and three-dimensions, obeying no-slip or stress-free boundaries conditions, satisfies $\max\,\text{Nu} \geq C' \text{Pe}^{2 / 3}/(\log  \text{Pe} )^{4/3}$ within logarithms of the \textit{a priori} upper bound $Nu \leq C\, \text{Pe}^{2/3}$.
Curiously, this lower bound coincides to a degree with the numerical results in the range of P\'eclet numbers explored: logarithmic differentiation yields
\begin{align}
\frac{d \log (\text{Nu} - 1)}{d \log \text{Pe} }  \approx \frac{2}{3} - \frac{4}{3} \frac{1}{\log \text{Pe}} \approx 0.55 \text{ at Pe} = 10^5 ,
\end{align}
not far from the  numerically computed $\text{Nu} \sim \text{Pe}^{0.54}$ result. Although at first we found no direct numerical evidence for the branching patterns constructed in \cite{tobasco2017} and \cite{tobasco2018}, a singular value decomposition revealed branching in its inception.  We wonder exactly how large $\text{Pe}$ must be for the higher modes to play a significant role in optimizing heat transport in a two-dimensional fluid layer.

%Such multi-scale structures appear to be necessary to achieve nearly optimal heat transport at higher $\text{Pe}$.
%Apparently, two-dimensional wall-to-wall optimal transport does not favor branching at moderate $\text{Pe}$. We find this to be rather mysterious, especially given the recent results of \cite{osakaheat2} where truly three-dimensional branching flows were found to achieve nearly globally optimal heat transport at significantly lower $\text{Pe}$.}

%This empirical observation suggests that most of the heat transport is indeed accounted for by a single tensor product mode.

The three-dimensional computations reported in \cite{osakaheat2} show a stark difference in flow structures and transport scaling as compared to the two-dimensional flows for both stress-free boundaries from \cite{HassanzadehChiniDoering2014} and no-slip boundaries from \sref{results} of the present paper (see also \cite{osakaheat}).
Three-dimensional versions of branching appear to achieve the optimal scaling $\text{Nu} \sim \text{Pe}^{2/3}$ already for $\text{Pe} \in [10^3, 10^4] $ with a prefactor within $10\%$ of the background upper bound computations from \cite{plasting_kerswell_2003}. We wonder if this $10\%$ difference is a manifestation of our proposed duality gap between the wall-to-wall problem and the background method bounds. In any case, the evidence is that optimal flows take advantage of the presence of a third spatial dimension to maximize thermal transport.

We conclude with a list of five fundamental questions remaining for the optimal wall-to-wall transport problem:
\begin{enumerate}
\item Are steady flows optimal?
\item Is it possible to prove the \textit{a priori} upper bound $\text{Nu} \leq C Pe^{2 / 3} / \log ( \text{Pe} )^{4/3}$ for two-dimensional flows as $\text{Pe} \rightarrow \infty$, and does the upper bound $\text{Nu} \leq C Pe^{6/11}$ hold for moderate $Pe$ instead?
\item Do there exist three-dimensional flow fields achieving $\text{Nu} \sim \text{Pe}^{2/3}$ as $\text{Pe} \rightarrow \infty$?
\item What do optimally transporting flows look like in other geometries, such as cylinders or domains with holes?
\item Do structural properties of optimal flows resemble those of buoyancy-driven steady and/or statistically stationary turbulent Rayleigh-B\'enard convection?
\end{enumerate}

\section{Acknowledgments}
We thank Rich Kerswell and Divakar Viswanath for many illuminating conversations. This work was supported in part by NSF Awards DMS-1344199 (ANS), DGE-0813964 and DMS-1812831 (IT), DMS-1515161 and DMS-1813003 (CRD), a Van Loo Postdoctoral Fellowship (IT) and a Guggenheim Foundation Fellowship (CRD).

 \appendix
 
 \section{Spatial Discretisation}
\label{disc_details}
In what follows we provide details of implementing spectral integration with regards to solving the Poisson and Stokes equation with respect to our boundary conditions and domain. 

 \subsection{Spectral Methods}
\label{spec}
Taking the Fourier transform of \eref{mpoisson} and \eref{mstokes} in the horizontal directions leads to the following set of ODE's to solve
\begin{align}
\left( D^2 - \beta_{n \ell}^2\right) \theta_{n\ell} &= f_{n\ell}
\end{align}
and
\begin{align}
\left(D^2 - \beta^2_{n \ell} \right)\mathbf{u}_{n\ell} &= \mathbf{f}_{n\ell} + Dp_{n \ell} \hat{z} + \imath  k_n p_{n \ell} \hat{x} + \imath \kappa_\ell p_{n \ell} \hat{y} \\
 \imath  k_{n} u_{n \ell} + \imath \kappa_\ell v_{n \ell} +Dw_{n \ell} &= 0
\end{align}
for $n, \ell \in \mathbb{Z}$, where $D$ the derivative in the vertical direction, and 
\begin{align}
k^2_{n\ell} &= (k_n )^2 + (\kappa_\ell)^2 \\
\beta^2_{n\ell} &= k^2_{n\ell} + c \\
k_n &= \frac{n \pi }{\Gamma_1} \\
\kappa_\ell &= \frac{\ell \pi}{\Gamma_2} \\
\imath &= \sqrt{-1}.
\end{align}
The square root denotes the principle branch.
Although we could discretise the spatial coordinates using Chebyshev matrices, we instead use spectral integration. This method of solving boundary value problems of the form
\begin{align}
(D-k)y = f
\end{align}
subject to boundary conditions has numerous advantages over the differentiation matrix approach as in \cite{Trefethen}. With spectral integration the operators that must be inverted have bounded condition numbers and are banded matrices as opposed to dense matrices with unbounded condition numbers. 

Instead of solving for functions on $z \in [0,1]$ it is more convenient to take $z \in [-1,1]$ and then convert results back to the original domain. Solutions in the $z\in [0,1]$ domain--denoted by subscripted $1$'s as in $\theta_1, \mathbf{u}_1$--are related to solutions in the $z \in [-1,1]$ domain--denoted by subscripted $2$'s as in $\theta_2, \mathbf{u}_2$--via the following relations
\begin{align}
\theta_1 &= \frac{1}{2}\theta_2 ,
\mathbf{u}_1 = 2 \mathbf{u}_2 ,
 \text{Pe}_1 = 4 \text{Pe}_2 ,
 (\text{Nu}-1)_1 =  (\text{Nu}-1)_2 ,
 \mu_1 = \mu_2 /16. 
\end{align}
When performing calculations we use the $[-1,1]$ domain but report results in terms of the original $z \in [0,1]$ domain. 

Computing averages (such as the Nusselt number) can be achieved with spectral accuracy. As mentioned in \cite{Boyd} the trapezoidal rule is spectrally accurate for periodic functions and for bounded domains there are quadrature weight formulas both in terms of the Chebyshev nodal and modal values. Furthermore all products are computed by taking the inverse transforms, multiplying in real space, and converting back to spectral space, as is common for pseudo-spectral methods. 

\subsection{Spectral Integration}
\label{spec_int}
We have seen that gradient ascent for the wall-to-wall problem reduces to solving differential equations of the form
\begin{align}
\label{building_block}
(D-k)y =f,
\end{align}
where $k\in \mathbf{R}$ and $z \in [-1,1]$. The differential equation has the solution
\begin{align}
y(z) &= C e^{kz} + e^{-kz} \int_{-1}^z e^{kx} f(x) dx,
\end{align}
where $C$ enforces boundary or integral constraints. This reduces the problem to quadrature and indeed the method that we adopt implicitly constructs the solution in this manner as noted in \cite{Viswanath2015159}.

To solve \eref{building_block} we use a modern form of spectral integration developed by Viswanath \cite{Viswanath2015159}. The general principle is remarkably simple. First compute the homogeneous solution
\begin{align}
\label{homog}
(D-k)y^h  = 0
\end{align}
and then the particular solution
\begin{align}
\label{partic}
(D-k)y^p = f
\end{align}
so that the general solution is then a linear combination of the particular and homogeneous solution
\begin{align}
\label{basic}
y = C y^h + y^p
\end{align}
where $C$ is a constant that enforces boundary conditions or integral constraints. This basic decomposition of the general solution of an ordinary differential equation serves as the primary building block in the construction of solutions to \pref{mpoisson} and \pref{mstokes}.

We now discuss how to construct $y^h$ and $y^p$ in the domain $z\in [-1,1]$. First write $y$ as Chebyshev expansion of the form
\begin{align}
y(z) = \frac{y_0}{2} P_0 + \sum_{n=1}^\infty y_n P_n(z)
\end{align}
where $P_n(x)$ $n = 0,1,2,...$ are the Chebyshev polynomials defined by 
\begin{align}
P_n(z) = \cos ( n \cos^{-1}(z)).
\end{align}
The factor of $1/2$ in front of the $y_0$ term is standard and convenient for formulas later on.

Let $\mathcal{T}_n(y)$ denote the $n$'th Chebyshev coefficent of $y$, e.g.
\begin{align}
\mathcal{T}_n ( y ) &= y_n \\
\label{anti}
\mathcal{T}_n \left( \int y \right) &= 
\begin{cases}
0 & \text{ for } n =0 \\
\frac{y_{n-1} - y_{n+1}}{2 n} & \text{ for } n > 0
\end{cases}
\end{align}
where $\int y$ here denotes a particular anti-derivative of $y$. The coefficients may be computed by the linear operator
\begin{align}
y_n &= \frac{2}{\pi}\int_{-1}^1 y(x) \frac{P_n(x)}{\sqrt{1-x^2}} dx.
\end{align}

Instead of working with \pref{building_block} directly we work with the equation in integral form 
\begin{align}
\label{build_int}
y - k \int y = \int f + C
\end{align}
where $C$ is a constant of integration. An important observation is that if $f = Dg$ for some function $g$, then we do not need to differentiate $g$ to write it in the form \pref{build_int} but rather we can directly use
\begin{align}
y - k \int y = g + C.
\end{align}
In other words we can avoid numerically differentiating $g$.

Use the relation \pref{anti} to construct \pref{build_int} as a system of equations for the Chebyshev coefficents, i.e.
\begin{align}
\mathcal{T}_n(y) - k \mathcal{T}_n \left(\int y \right) &= \mathcal{T}_n\left( \int f \right) +  \mathcal{T}_n\left( C \right) \\
&\Rightarrow \nonumber \\ 
y_0  &=  2C  \text{ for } n =0 \\
y_n -k \frac{y_{n-1} - y_{n+1}}{2 n} &=  \frac{f_{n-1} - f_{n+1}}{2 n} \text{ for } n > 0
\end{align}
The $f_n$ are the Chebyshev coefficents of the forcing function $f$. Note that any choice of $C$ yields a particular solution to the problem, but does not enforce the proper boundary conditions. This is an infinite dimensional tridiagonal system of equations for the Chebyshev coefficients of $y$.

 In order to be amenable to computation any such system of equations must be truncated. Thus we assume that the solution $y$ and forcing function $f$ is well represented by a finite truncation
 \begin{align}
 y(z) &= \frac{y_0}{2} + \sum_{n=1}^{N-2} y_n P_n(z) + \frac{y_{N-1}}{2} P_{N-1}(z) \\
  f(z) &= \frac{f_0}{2} +  \sum_{n=1}^{N-2} f_n P_n(z) + \frac{f_{N-1}}{2} P_{N-1}(z).
 \end{align}
 For example, with $N=6$ we would have the following system of equations to solve,
 \begin{align}
 \begin{bmatrix}
1 & 0 & 0 & 0 & 0 &0 &  \\
-\frac{k}{2} & 1 & \frac{k}{2} & 0 & 0 & 0  \\
0 & -\frac{k}{2 \cdot 2} & 1 & \frac{k}{2 \cdot 2} & 0 & 0 \\
0 & 0 & -\frac{k}{2 \cdot 3} & 1 & \frac{k}{2 \cdot 3} & 0 \\
0 & 0 & 0 &-\frac{k}{2 \cdot 4} & 1 & 0  \\
0 & 0 & 0 & 0 & -\frac{k}{2 \cdot 5} & 1
\end{bmatrix}
\begin{bmatrix}
y_0 \\
y_1 \\
y_2 \\
y_3 \\
y_4 \\
y_5 
\end{bmatrix}
&= 
\begin{bmatrix}
2 C \\
 \frac{f_0 -f_2}{2} \\
\frac{f_1 - f_3}{2\cdot 2} \\
\frac{f_2 - f_4}{2 \cdot 3} \\
\frac{f_3 - f_5}{2 \cdot 4} \\
\frac{f_4 }{2 \cdot 5} 
\end{bmatrix}.
 \end{align}
 The use of this finite representation has additional benefits. The Chebyshev spectral coefficients of $y$  are related to the nodal values of $y$ evaluated at $\cos( \pi j/(N-1))$ for $j=0,1,..., N-1$ via the fast-cosine transform. This allows us to quickly convert spectral coefficients into real space, and herein lies one of the many advantages of using a Chebyshev series.

We are now in a position to show how to construct numerical homogeneous and particular solutions $y^h$ and $y^p$. To construct $y^h$ we solve the following problem
\begin{align}
\label{aux}
(D-k) v &= \frac{k}{2}
\end{align}
subject to $\mathcal{T}_0 (v) = 0$. 
The general solution to \pref{aux} is 
\begin{align}
v &= C y^h + v^p \\
v^p &= -1/2 \\
y^h(z) &= e^{k z}
\end{align}
where  $v^p = -1/2$ is the particular solution. The condition $\mathcal{T}_0(v) = 0$ guarantees that $C \neq 0$ since
\begin{align}
C \mathcal{T}_0( y^h ) &= \mathcal{T}_0( v - v^p) = \mathcal{T}_0( v ) - \mathcal{T}_0 ( v^p)  = 0 - (-1) = 1
\end{align}
and $\mathcal{T}_0( v^h ) \neq 0$. Thus the homogeneous solution is $y^h = v + 1/2$. Denoting the Chebyshev series of $v$ by $v_n$ for $n=0,...,N-1$, we find $v$ by solving the system of equations
\begin{align}
v_0 &= v_{N-1} = 0 \\
v_n -k \frac{v_{n-1} - v_{n+1}}{2 n} &=  \frac{f_{n-1} - f_{n+1}}{2 n} \text{ for } 0< n < N-1.
\end{align}
We set the $N-1$'st Chebyshev coefficent of $v$ and $f$ as zero for convenience. For example, for $N=6$ we solve the system of equations,
\begin{align}
 \begin{bmatrix}
1 & 0 & 0 & 0 & 0 &0 &  \\
-\frac{k}{2} & 1 & \frac{k}{2} & 0 & 0 & 0  \\
0 & -\frac{k}{2 \cdot 2} & 1 & \frac{k}{2 \cdot 2} & 0 & 0 \\
0 & 0 & -\frac{k}{2 \cdot 3} & 1 & \frac{k}{2 \cdot 3} & 0 \\
0 & 0 & 0 &-\frac{k}{2 \cdot 4} & 1 & 0  \\
0 & 0 & 0 & 0 & 0 & 1
\end{bmatrix}
\begin{bmatrix}
v_0 \\
v_1 \\
v_2 \\
v_3 \\
v_4 \\
v_5 
\end{bmatrix}
&= 
\begin{bmatrix}
0 \\
-\frac{k}{2}\\
0 \\
0 \\
0 \\
0 
\end{bmatrix}
\end{align}
for the Chebyshev coefficients of $v$ and then add $1$ to $v_0$ to construct $y^h$.

For the construction of the particular solution $y^p$ we solve \pref{partic} subject to $\mathcal{T}_0(y^p) = 0$. Thus we solve the system of equations
\begin{align}
v_0 &= v_{N-1} = 0 \\
v_n -k \frac{v_{n-1} - v_{n+1}}{2 n} &=  \frac{f_{n-1} - f_{n+1}}{2 n} \text{ for } 0< n < N-1.
\end{align}
For $N=6$ implies solving the tridiagonal system
 \begin{align}
 \begin{bmatrix}
1 & 0 & 0 & 0 & 0 &0 &  \\
-\frac{k}{2} & 1 & \frac{k}{2} & 0 & 0 & 0  \\
0 & -\frac{k}{2 \cdot 2} & 1 & \frac{k}{2 \cdot 2} & 0 & 0 \\
0 & 0 & -\frac{k}{2 \cdot 3} & 1 & \frac{k}{2 \cdot 3} & 0 \\
0 & 0 & 0 &-\frac{k}{2 \cdot 4} & 1 & 0  \\
0 & 0 & 0 & 0 & 0 & 1
\end{bmatrix}
\begin{bmatrix}
y_0^p \\
y_1^p\\
y_2^p \\
y_3^p \\
y_4^p \\
y_5^p 
\end{bmatrix}
&= 
\begin{bmatrix}
0 \\
 \frac{f_0 -f_2}{2} \\
\frac{f_1 - f_3}{2\cdot 2} \\
\frac{f_2 - f_4}{2 \cdot 3} \\
\frac{f_3 - f_5}{2 \cdot 4} \\
0
\end{bmatrix}.
 \end{align}
 
Now that we have constructed the homogeneous and particular solutions we can enforce boundary conditions. Given $y(a) = b$ for $a \in [-1,1]$ the value of the constant $C$ is determined:
\begin{align}
y(a) = C y^{h}(a) + y^p(a) \text{ } \Rightarrow C = \frac{b-y^p(a)}{y^h(a)}.
\end{align}
Typically we enforce the boundary conditions at the endpoint $z=\pm 1$ for which we have readily available formulas to compute the values of $y^h$ and $y^p$ in terms of their Chebyshev coefficents. For $f = f_0 /2 +\sum_{n=1}^{N-2} f_n P_n + P_{N-1}f_{N-1}/2$ we compute
\begin{align}
\label{end1}
f(-1) &= f_0/2 + \sum_{j=1}^{N-2} (-1)^j f_j + (-1)^{N-1} f_{N-1}/2 \\
\label{end2}
f(1) &= f_0/2 + \sum_{j=1}^{N-2}  f_j + f_{N-1}/2 \\
\label{deriv1}
f'(-1) &= \sum_{j=1}^{N-2} (-1)^{j+1} j^2 f_j + (-1)^{N} (N-1)^2 f_{N-1}/2 \\
\label{deriv2}
f'(1) &= \sum_{j=1}^{N-2} j^2 f_j +  (N-1)^2 f_{N-1}/2 \\
f''(1) &= \frac{1}{3}\sum_{j=2}^{N-2} (j^4-j^2) f_j +  \left[ (N-1)^4 - (N-1)^2 \right] f_{N-1}/2 \\
f''(-1) &= \frac{1}{3}\sum_{j=2}^{N-2} (-1)^{j}  (j^4-j^2)  f_j + (-1)^{N-1}  \left[ (N-1)^4 - (N-1)^2 \right] f_{N-1}/2. 
\end{align}

We now show how to construct solutions to the second order equation
\begin{align}
(D^2 - k^2)y = f.
\end{align}
For this problem we have two homogeneous solutions and one particular solution so that the general solution to this differential equation is of the form
\begin{align}
y &= C_1 y^{h_1} + C_2 y^{h_2} + y^p.
\end{align}
We solve this as a system of two different equations
\begin{align}
(D-k)v &= f \\
(D+k)y &= v.
\end{align}
First we find the particular and homogeneous solution to 
\begin{align}
(D-k)v &= f 
\end{align}
as described previously so that we have
\begin{align}
v =C v^{h_1} + v^p.
\end{align}
Then we find $y^{h_2}$ and $y^p$ by solving
\begin{align}
(D+k) y^{h_2} &= y^{h_1} \\
(D+k) y^p &= v^p 
\end{align}
subject to $\mathcal{T}_0(y^{h_2}) = 0$ and $\mathcal{T}_0(y^p) = 0$ via spectral integration for the particular solution. Note that $y^{h_2}$ constructed in this manner is linearly independent from $y^{h_1}$. To enforce boundary conditions we must now invert a matrix for the coefficients $C_1$ and $C_2$. For example for boundary condition enforced at the endpoints $z=\pm$ we have
\begin{align}
\begin{bmatrix}
y^{h1}(-1) & y^{h2}(-1) \\
y^{h1}(1) & y^{h2}(1)
\end{bmatrix}
\begin{bmatrix}
C_1 \\
C_2
\end{bmatrix}
&= 
\begin{bmatrix}
y(-1) - y^p(-1) \\
y(1) - y^p(1)
\end{bmatrix}
\end{align}
for the coefficients $a$ and $b$. 

 Once we have our solution $y$ we find derivatives without recourse to differentiation matrices or Fourier transform methods.  Indeed, in the first-order case, i.e. for $y$ that satisfy
\begin{align}
(D-k)y = f,
\end{align}
we find the derivative by simply rearranging the equation
\begin{align}
Dy = f + k y.
\end{align}
Hence differentiating is obtained by merely summing the the forcing function and the solution multiplied by $k$. For the second order case, i.e. for $y$ that satisfy
\begin{align}
C_1 y^{h_1} + C_2 y^{h_2} + y^p &= y \\
(D^2 - k^2)y &= f \\
(D-k) y^{h_1} &= 0 \\
(D+k) y^{h_2} &= y^{h_1} \\
(D-k) y^{dp} &= f \\
(D+k)y^p &= y^{dp},
\end{align}
we compute
\begin{align}
\label{deriv_formula}
D y &= D( C_1 y^{h_1} + C_2 y^{h_2} + y^p)  \\
&= k C_1 y^{h_1} + C_2 ( y^{h_1} - k y^{h_2}) - k y^p + y^{dp} \\
D^2 y &= k^2 y + f.
\end{align}
It seems that computing derivatives in this manner lead to an order of magnitude improvement of the relative error over other methods (either differentiation matrices or Fourier methods). Using these formulas we compute the derivatives of $y$ at the endpoints in an alternative manner. Instead of using \pref{deriv1} and \pref{deriv2} to the solution $y$, we apply \pref{end1} and \pref{end1} to \pref{deriv_formula}. From numerical experimentation it does not seem to matter which way the derivatives were evaluated at the endpoint.

We now have all the pieces to solve the modified Poisson's equation \pref{mpoisson}. As was the case with the Chebyshev series we must truncate the number of horizontal Fourier wave-modes in hopes that we achieve a good representation of our solution. This problem is linear thus we solve
\begin{align}
(D^2-\beta^2_{n\ell} ) \theta_{n\ell} = f_{n\ell}
\end{align}
subject to $\theta_{n\ell} ( z = \pm 1) = 0$, mode by mode. Here
\begin{align}
k^2_{n\ell} &= (k_n)^2 + (\kappa_\ell)^2 \\
\beta^2_{n\ell} &= k^2_{n\ell} + c \\
c &\geq 0 \\
k_n &= \frac{n \pi }{\Gamma_1} \text{ for } n=  -N/2+1, ..., 0,..., N/2 \\
\kappa_\ell &= \frac{\ell \pi}{\Gamma_2} \text{ for } \ell = -L/2+1, ..., 0,..., L/2.
\end{align}
and $N, L$ are even. Thus we solve $N\times L$  second order boundary value problems. By using symmetry or realness of the variables we reduce computation. The $n = \ell =0$ mode for $c=0$ must be handled separately depending on boundary conditions. For example, if $\partial_z \theta_{00} (z = \pm 1 )=0$ there is the solvability requirement that $0=\int_{z=-1}^1 f_{00}(z) dz$ otherwise no solution exists. Changing the value of $\theta_{00}$ by any constant still produces a solution, thus instead we replace one of the boundary conditions with a normalization condition such as $\int_{-1}^1 \theta_{00} dz = 0$. 

 Solving the modified Stokes equation \pref{mstokes} is more complicated and is the subject of the next section.

\subsection{Kleiser-Schumann Algorithm}
\label{kl_sch}
The Kleiser-Schumann algorithm is a method for solving the modified Stokes problem \pref{mstokes}, see \cite{ksalg,Viswanath2013414}. Since the modified Stokes problem is linear we solve wave-number by wave-number equations of the form
\begin{align}
(D^2 - \beta^2_{n\ell}) \mathbf{u}_{n\ell} &= \mathbf{f}_{n\ell} - \hat{\nabla} p_{n\ell} \\
\hat{\nabla} \cdot \mathbf{u}_{n\ell} &= 0 
\end{align}
where
\begin{align}
k^2_{n\ell} &= (k_n )^2 + (\kappa_\ell)^2 \\
\beta^2_{n\ell} &= k^2_{n\ell} + c \\
\hat{\nabla} &= \imath k_n \hat{x} + \imath \kappa_\ell \hat{y} + D \hat{z} \\
k_n &= \frac{n \pi }{\Gamma_1} \text{ for } n=  -N/2+1, ..., 0,..., N/2 \\
\kappa_\ell &= \frac{\ell \pi}{\Gamma_2} \text{ for } \ell = -L/2+1, ..., 0,..., L/2 \\
\mathbf{f}_{n\ell} &= f^1_{n\ell} \hat{x} + f^2_{n\ell} \hat{y} + f^3_{n\ell} \hat{z} \\
\mathbf{u}_{n\ell} &= u_{n\ell} \hat{x} + v_{n\ell} \hat{y} + w_{n\ell} \hat{z}.
\end{align}
Here $\imath = \sqrt{-1}$ where the square-root is interpreted as the principle branch.
From now on we drop the subscript $n\ell$ with the understanding that the modified Stokes problem must be solved for all values of $n$ and $\ell$ individually. For now we consider the $k \neq 0$ case and discuss how to handle this mode separately at the end. 

The single wave-number problem is to solve 
\begin{align}
(D^2 - \beta^2) \mathbf{u} &= \mathbf{f}- \hat{\nabla} p \\
\hat{\nabla} \cdot \mathbf{u} &= 0 .
\end{align}
Taking the divergence of the first equation yields the following equation for $p$
\begin{align}
(D^2 - k^2) p &= \hat{\nabla} \cdot \mathbf{f}.
\end{align}
As we have seen before we may write the general solution to this problem as the sum of two homogeneous terms and a particular solution
\begin{align}
p = C_1 p^{h_1} + C_2 p^{h_2} + p^p.
\end{align}
Hence we may write the equation for $\mathbf{u}$ as
\begin{align}
(D^2 - \beta^2) \mathbf{u} &= \mathbf{f}-\hat{\nabla} p^p - C_1 \hat{\nabla}p^{h_1}  - C_2 \hat{\nabla}p^{h_2} .
\end{align}
We split the problem of finding solutions to $\mathbf{u}$ into three parts. Let $\mathbf{u}\hspace{1mm}^i$ for $i=1,2,3$  be solutions to
\begin{align}
(D^2 - \beta^2) \mathbf{u}\hspace{1mm}^{1} &= \hat{\nabla}p^{h_1}  \\
(D^2 - \beta^2) \mathbf{u}\hspace{1mm}^{2} &= \hat{\nabla}p^{h_2} \\
(D^2 - \beta^2) \mathbf{u}\hspace{1mm}^{3} &= \mathbf{f}-\hat{\nabla} p^p   .
\end{align}
where each $\mathbf{u}\hspace{1mm}^i$ for $i=1,2,3$ satisfies the boundary conditions for $\mathbf{u}$.
With this we write $\mathbf{u}$ as 
\begin{align}
\label{vec_sum}
\mathbf{u} &= \mathbf{u}\hspace{1mm}^3 - C_1 \mathbf{u}\hspace{1mm}^1 - C_2\mathbf{u}\hspace{1mm}^2.
\end{align}
To find $C_1$ and $C_2$ we use auxiliary conditions derived by enforcing incompressibility on the boundary.

For no-slip boundary conditions we have $\partial_z w (z = \pm 1 ) = 0$ and for stress-free boundary conditions $\partial_{zz} w( z =\pm 1)  = 0$. That is to say \pref{vec_sum} applies to each component hence to the vertical velocity $w$
\begin{align}
w &= w^3 - C_1 w^1 - C_2 w^2.
\end{align}
For example with no-slip boundary conditions we apply the vertical derivative $D$ to both sides and solve the following system of equations for $C_1$ and $C_2$
\begin{align}
\begin{bmatrix}
Dw^1(z =1) & D w^2 (z=1) \\
Dw^1(z =-1) & D w^2 (z=-1) 
\end{bmatrix}
\begin{bmatrix}
C_1 \\
C_2
\end{bmatrix}
&=
\begin{bmatrix}
Dw^3(z= 1)  \\
Dw^3 ( z = -1 ) 
\end{bmatrix}
\end{align}
For fixed time-step sizes the matrix for are precomputed and factorized only once.  Furthermore $\mathbf{u}\hspace{1mm}^1$ and $\mathbf{u}\hspace{1mm}^2$ is precomputed. Hence at each time step we only need to solve for $\mathbf{u}\hspace{1mm}^3$ and the coefficients $C_1$ and $C_2$. 

For this problem the $k=0$ case must be handled separately as well. This is due to the fact that the homogeneous solution for the pressure is of the form $ p^{h_1} = 1/2$ and $p^{h_2} = z/2$. Letting $\mathbf{u}\hspace{1mm}^1 = (u^1, v^1, w^1)$ the $k=0$ case implies that $\mathbf{u}\hspace{1mm}^1 = 0$ for  no-slip boundary conditions and stress-free boundary conditions. For $\beta =0$ one has $(u^1,v^1,w^1) = (D_1, D_2, 0)$ in the stress-free case where $D_1$ and $D_2$ are arbitrary constants. By specifying that $(u^1,v^1)$ are mean zero we may set these arbitrary constants to zero. The constant $C_1$ becomes a free parameter that doesn't affect the physical flow field $\mathbf{u}$. This corresponds to the fact that the pressure may be changed by an arbitrary constant. We may choose this value such that the average of $p$ is zero, but this is by no means necessary. Now we must find $C_2$. From incompressibility and from the boundary conditions we see that $Dw = 0 \Rightarrow w = 0$ which implies\footnote{Note that for $\beta = 0$ we have $w^2  = (z^2-1)/4$ and for $\beta \neq 0$ we have $w^2 = -\frac{1}{2\beta^2}+\frac{\sinh(\beta) \cosh(\beta z)}{\beta^2 \sinh (2 \beta)}$.} $C_2 w^2 = w^3$. Taking the derivative we find that $Dw^3(z \pm 1) / Dw^2(z \pm 1) = C_2$. This may appear overconstrained, but since we are guaranteed that that the functions $w^2$ and $w^3$ are proportional to one another we could take either conditions to evaluate the constant $C_2$.

\section{Optimality Condition for Domain Size}
\label{conv_crit_domain}
Prior work \citep{HassanzadehChiniDoering2014} indicates that there seems to be an optimal domain size in the periodic direction $x \in [0,\Gamma]$.
The derivative of the functional $\mathcal{F}$ with respect to domain size $\Gamma$, is easiest to compute by rewriting  $\mathcal{F} = \langle \mathcal{L} \rangle$ with 
\begin{align}
 \mathcal{L} = u_3 \theta + \varphi \mathbf{u}\cdot \nabla \theta - \nabla \varphi \cdot \nabla \theta + \varphi u_3  + \mu\left(\text{Pe}^2 -  | \nabla \mathbf{u} |^2\right) + \nabla p \cdot \textbf{u} 
\end{align}
Differentiating this functional with respect to $\Gamma$ evaluated on a solution to the Euler-Lagrange equations yields
\begin{align}
\frac{\delta \mathcal{F}}{\delta \Gamma} &=  - \frac{1}{\Gamma}\langle \mathcal{F} \rangle + \frac{1}{\Gamma}\int_0^1 \mathcal{H}^x dz \\
-\mathcal{H}^x &= -\mathcal{L} - \partial_x \theta \varphi u_1 + \mu |\partial_x \mathbf{u} |^2 - p \partial_x u + 2\partial_x \theta \partial_x \phi 
\end{align}
The latter part is the spatial Hamiltonian density of $\mathcal{L}$.
It may seem that $\frac{\delta \mathcal{F}}{\delta \Gamma} $ is a function of $x$, but the Lagrangian $\mathcal{L}$ has no explicit spatial dependence hence the quantity $\int_0^1\mathcal{H}^x dz$ is independent of the horizontal variable $x$.
In other words $\int_0^1\mathcal{H}^x dz = \langle \mathcal{H}^x \rangle$.
This allows us to simplify the formula and compute\footnote{If this calculation is unfamiliar to the reader we refer to \cite{SouzaThesis2016} but the essence of the calculation comes from the first exercise of Feynman and Hibbs \cite{Feynman}.}
\begin{align}
\frac{\delta \mathcal{F}}{\delta \Gamma} &= \frac{1}{\Gamma}  \langle \partial_x \theta \varphi u_1 + \mu |\partial_x \mathbf{u} |^2 - p \partial_x u + 2\partial_x \theta \partial_x \phi \rangle 
\end{align}

The optimal domain size condition for $\mathcal{S}$ is similar but not exactly the same due to a redefinition of the pressure term.

We include an aspect ratio flow of the form $\partial_\tau \Gamma = \frac{\delta \mathcal{F}}{\delta \Gamma}$ in addition to the previous time-stepping procedures.
There are subtleties associated with implementing this flow to make the procedure successful.
Empirically it is found to be useful to withhold evolution of the aspect ratio until one was ``close" to a solution of the Euler-Lagrange equations. 
Furthermore taking a time-step in tandem with the other evolution seemed prohibitively slow, thus the evolution is only taken every $n$'th time-step with respect to the other evolutions, where $n=5$ works well for our purposes.
The overhead of changing the domain slows down computations by a negligible amount once these modifications are implemented.

\newpage

\bibliographystyle{jfm}
\bibliography{mybibfile}

\end{document}